\documentclass[aps,prd,showpacs, superscriptaddress,  nofootinbib, floatfix, onecolumn,  amsmath,11pt, tightenlines]{article}%
\usepackage{bbm}
\usepackage[numbers,sort&compress]{natbib}
\usepackage{graphicx}
\usepackage{amsmath}
\usepackage{amssymb}
\usepackage{hyperref}
\usepackage{indentfirst}
\usepackage{bbm}
\usepackage{mathrsfs}
\def\be{\begin{equation}}
\def\ee{\end{equation}}
\def\bea{\begin{eqnarray}}
\def\eea{\end{eqnarray}}
\def\bsp{\be\begin{split}}

\def\bes{\be  \begin{split}}

 \makeatletter

\newcommand{\Rmnum}[1]{\expandafter\@slowromancap\romannumeral #1@}
    \newcommand{\ba}{\begin{eqnarray}}
    \newcommand{\ea}{\end{eqnarray}}
    \newcommand{\nbl}{\overleftarrow{D}}
    \newcommand{\nbr}{\overrightarrow{D}}
    \newcommand{\nblr}{\overleftrightarrow{D}}
    \newcommand{\nl}{\overleftarrow{\partial}}
    \newcommand{\nr}{\overrightarrow{\partial}}
    \def\eps{\varepsilon_{ijk}}

    \makeatother

\renewcommand{\title}[1]{\vbox{\center\LARGE{#1}}\vspace{5mm}}
\renewcommand{\author}[1]{\vbox{\center\large{#1}}\vspace{5mm}}
\newcommand{\address}[1]{\vbox{\center\em#1}}

\begin{document}

\begin{titlepage}
\begin{flushright}
UK/11-09
\end{flushright}

\title{\vspace{1.0in} {\bf Is $1^{-+}$ Meson a Hybrid?}}

\author{Yi-Bo Yang$^{ab}$, Ying Chen$^{ab}$, Gang Li$^{c}$, and Keh-Fei Liu$^{d}$}

\address{\it
$^{a}$Institute of High Energy Physics, Chinese Academy of Sciences, Beijing 100049, P.R. China \\
$^{b}$~Theoretical Center for Science Facilities, Chinese Academy of Sciences, Beijing 100049,
 P.R. China\\
$^{c}$Department of Physics, Qufu Normal University, Qufu, 273165, P.R. China\\
$^{d}$Dept.\ of Physics and Astronomy, University of Kentucky, Lexington, KY 40506, USA \\
}

\vskip 0.5cm 

\begin{abstract}
   We calculate the vacuum to meson matrix elements of the dimension-4 operator $\bar{\psi}\gamma_4\nblr_i \psi$
and dimension-5 operator $\bar{\psi}\eps\gamma_j\psi B_k$ of the $1^{-+}$ meson on the lattice and compare them to the corresponding matrix elements of the ordinary mesons to discern if it is a hybrid.
For the charmoniums and strange quarkoniums, we find that the matrix elements of $1^{-+}$ are 
comparable in size as compared to other known $q\bar{q}$ mesons. They are particularly similar
 to those of the $2^{++}$ meson, since their dimension-4 operators are in the same Lorentz multiplet.
  Based on these observations, we find no evidence to support the
notion that the lowest $1^{-+}$ mesons in the $c\bar{c}$ and $s\bar{s}$ regions are hybrids.
As for the exotic quantum number is concerned, the non-relativistic reduction reveals that the leading terms in the dimension-4 and dimension-5
operators of $1^{-+}$ are identical up to a proportional constant and it involves a center-of-mass momentum operator of the quark-antiquark pair.
This explains why $1^{-+}$ is an exotic quantum number in the constituent quark model where the center of mass of the
$q\bar{q}$ is not a dynamical degree of freedom. Since QCD has gluon fields in the context of the flux-tube which is appropriate for heavy quarkoniums to allow the valence
$q\bar{q}$ to recoil against them, it can accommodate such states as $1^{-+}$. By the same token, hadronic models with additional constituents besides the quarks can also accommodate the $q\bar{q}$ center-of-mass motion.
To account for the quantum numbers of these $q\bar{q}$ mesons in QCD and hadron models in the non-relativistic case, the parity and total angular momentum should be modified to
$P = (-)^{L + l +1}$ and
$\vec{J} = \vec{L} + \vec{l} + \vec{S}$, where $L$ is the orbital angular momentum of the $q\bar{q}$ pair in the meson.
\end{abstract}

 \end{titlepage}

%
%
%
%
%
%

\vfill

\section{Introduction}

     In the course of studying the glueball spectrum in the MIT bag model~\cite{jj76,djl81,bcm82,chp83} and potential
models~\cite{bar81,cs82}, it is an underline assumption that there are valence gluons as are quarks. It is then a natural extension
to consider hybrids of constituent quarks and gluons in the form of $q\bar{q}g$. This has been studied in
the potential models~\cite{hm78,lw80}, bag model~\cite{bar79,bc82,cs83,bcv83,fps84}, flux-tube model~\cite{ip85,bcs95}, QCD
sum rules~\cite{lnp84,cn00,Huang:2011nv}, \mbox{ADS/QCD~\cite{kk09}} and lattice QCD~\cite{lmb97,bhd97,jkm99,lm02,bbg03,ml03,hlw04,dep09,dud11}.
One of the interesting attributes of these hybrids is that they can have exotic $J^{PC}$ quantum numbers -- these are
$J^{PC}$'s that are not accessible by the $q\bar{q}$ mesons in the constitute quark model where
the charge and parity of a $q\bar{q}$ meson are given by
\bea   \label{PC}
P & =& (-)^{l+1} \nonumber \\
C &=& (-)^{l+S},
\eea
and the angular momentum by
\be    \label{J}
\vec{J} = \vec{l} + \vec{S}.
\ee
 In light of this, these hybrids with
exotic quantum numbers, particularly the $1^{-+}$ has been studied in the above quoted references. Experimentally, there are two candidates for the $1^{-+}$ --- one is $\pi_1 (1400)$~\cite{ald88} and the other is $\pi_1 (1600)$~\cite{ada98}. They are observed in the $\eta \pi$ and $\rho \pi$ channels.

In view of fact that exotic quantum numbers are not accessible by the constituent quark-antiquark pair, it
is suggested that the interpolation field for the hybrids of the $q\bar{q}g$ type will necessarily involve a gauge field
tensor, i.e. of the form $\bar{\psi} \Theta \psi G$ where $\Theta$ involves $\gamma$ matrices and covariant derivatives and
$G$ stands for the field tensor $G_{\mu\nu}$. It is an operator with dimension $\ge$ 5. However, it was pointed out
by B.A. Li more than 30 years ago that these exotic quantum number can be constructed from the quark bilinears $\bar{\psi} \Theta \psi$ without the field tensor~\cite{li75}. For example, the $J^{PC}$ of $\bar{\psi} \gamma_4 \nblr_i \psi$ is $1^{-+}$ which is a dimension-4 operator. This type of operators have been constructed on the lattice~\cite{lm02} and lattice calculations have been calculated with them in addition to the dimension-5 operator $\eps \bar{\psi}\gamma_j\psi B_k$~\cite{lm02,dep09,dud11}. The exotic mesons can be in the form of tetraquark mesoniums $qq\bar{q}\bar{q}$ which will require a dimension 6 interpolation field. We will not address them in the present work.

 The existence of the dimension-4 operator for $1^{-+}$ that does not involve the gauge filed tensor raises several questions:
\begin{itemize}

\item  Since there exists an interpolation operator which does not involve the field tensor, does that mean the meson
       with this interpolation field is not an hybrid? One could point out that the dimension-4 operators involve a covariant
       derivative which allows it to couple to a constituent gluon, unlike the dimension-3 operators $\bar{\psi} \Gamma \psi$,
       where $\Gamma$ is a $\gamma$ matrix, for the pseudoscalar and vector mesons. However, one can counter this argument by
       pointing out that the tensor meson ($2^{++}$), like $1^{-+}$, does not have dimension-3 interpolation field. The minimum dimension of its interpolation field is a dimension-4 operator
$\bar{\psi} \gamma_i \nblr_j \psi$ which is very similar to that of the dimension-4
       operator for $1^{-+}$~\cite{li75} and yet $2^{++}$ is an ordinary quantum number. So, how does one find out whether a
       meson is a hybrid or not?

\item  Since $1^{-+}$ is an exotic quantum number, how come one can have an operator which does not involve the field
       tensor? If one carries out a non-relativistic reduction of the operator, would one be able to reveal why it is
       not accessible to the constituent quark model?

\end{itemize}

   To answer these questions, we shall establish criteria for identifying the hybrid and carry out a lattice calculation with both the dimension-4
and dimension-5 interpolation fields to compare their respective spectral weights against those of ordinary mesons. We
will also carry out a non-relativistic reduction to figure out why the exotic quantum numbers are not accessible to
the constituent quark model. We shall present the meson interpolation fields organized in dimensions 3, 4, and 5 for
various mesons in Sec. II, set criteria for distinguishing hybrids from ordinary mesons, and discuss the origin of the exotic quantum
numbers.  The numerical details are given in Sec. III and the results are given in Sec. IV. We will end with a summary in Sec. V.

\section{Formalism}\label{sec:theory}

      We shall discuss several types of meson interpolation fields and set up criteria in order to distinguish
the hybrids from the ordinary mesons via the vacuum-to-meson transition matrix elements.

\subsection{Meson interpolations fields and criteria for hybrids}  \label{sec:interpolation}

    In lattice calculations, one relies on interpolation fields with the desired quantum numbers (e.g. $J^{PC}$, isospin,
strangeness, etc.) to project to the physical spectrum with the corresponding quantum numbers. In the following we give a
list of these interpolation fields for the low-lying ordinary mesons (pseudoscalar, vector, axial-vector, scalar and
tensor) and  $1^{-+}$. They are classified according to the following types:
\begin{itemize}
\item  $\bar{\psi} \Gamma \psi$ ($\Gamma$ is a gamma matrix), a dimension-3 operator, is labeled as the $\Gamma$-type;
\item $\bar{\psi} \Gamma \times \nblr\psi$ ($\nblr = \nbr - \nbl$), a dimension-4 operator, is labeled as the D-type;
\item $\bar{\psi} \Gamma \times B \psi$ ($B_i\equiv\frac{1}{2}\eps G_{jk}$), a
dimension-5 operator, is labeled as the B-type.
\end{itemize}

 A more complete list can be found in Ref.~\cite{lm02}.

 \begin{table}[htb]
 \caption{Interpolation operators $\bar{\psi}\Gamma\psi$ (dimension 3, $\Gamma$-type), $\bar{\psi} \Gamma \times \protect\nblr\psi$
(dimension 4, \mbox{D-type}), and $\bar{\psi} \Gamma \times B \psi$ (dimension 5, B-type). $\Sigma_i\equiv\frac{1}{2}\protect\eps\sigma_{jk}$
and repeated indices are summed over.} \label{tab:Operators}
 \begin{center}
 \begin{tabular}{c|c|c|c}
\hline
            & $\Gamma$   & $D$     &  $B$ \\
\hline
$0^{-+}$  &$\gamma_5$       &$\Sigma_i \nblr_i$     &$\gamma_i B_i$ \\
$1^{--}$  &$\gamma_i$       &$\nblr_i$              &$\gamma5B_i$   \\
$0^{++}$  &$\mathbb{I}$     &$\gamma_i \nblr_i$     &$\Sigma_i B_i$ \\
$1^{++}$  &$\gamma_5\gamma_i$&$\eps\gamma_j\nblr_k$ &$\eps\Sigma_jB_k$\\
$1^{+-}$  &$\Sigma_i        $&$\gamma_5 \nblr_i$        &$B_i$          \\
$2^{++}$  &                 &$|\eps|\gamma_j\nblr_k$&$|\eps|\Sigma_jB_k$\\
$1^{-+}$  &                 & $\gamma_4\nblr_i$     & $\eps\gamma_jB_k$\\
          &                 &$ \eps\Sigma_j\nblr_k$    & \\
\hline
\end{tabular}
 \end{center}
 \end{table}

   Here, we only list $1^{-+}$ as an example of mesons with exotic quantum numbers that cannot be accessed by dimension-3 operators. We should point out that ordinary $J=2$ mesons
do not have dimension-3 interpolation operators either.
 There are two kinds of dimension-4 $1^{-+}$ operators
($\bar{\psi}^a\gamma_4\nblr_i \psi^a$ and $\eps\bar{\psi}^a\Sigma_j\nblr_k\psi^a$). These two kinds of operators have
very similar non-relativistic forms as will be discussed in Sec. \ref{subsec:non-rela}.

A meson correlator at zero momentum is
\be
C_{ij}(t) = \sum_{\vec{x}} \langle \mathcal{O}_i(\vec{x}, t) \mathcal{O}_j(0,0)\rangle.
\ee
At large time separation, it is dominated by the lowest state of the spectrum with the prescribed quantum number
\be
C_{ij}(t)_{\stackrel{-\longrightarrow}{t \rightarrow \gg a}} \frac{1}{2 m} \langle 0| \mathcal{O}_i|M\rangle \langle M| \mathcal{O}_j|0\rangle e^{-mt}
\ee
where $m$ is the mass of the lowest state. Besides the mass, one also obtains the vacuum to meson transition matrix
elements $ \langle 0| \mathcal{O}|M\rangle$.

We should point out that, notwithstanding claims in many lattice calculations, the interpolation operators do not
necessarily reflect the nature of the composition of the hadrons. They merely reveal how strongly the operators couple to the specific hadron,
such as realized in decay constants. For example, the topological charge operator $G_{\mu\nu}\tilde{G}_{\mu\nu}$ projects to $\eta$ and
$\eta'$ strongly.
From the anomalous Ward identity for massless fermions $\partial_{\mu} A_{\mu}^0 =  \frac{N_f}{16\pi^2} G_{\mu\nu}\tilde{G}_{\mu\nu}$, one has
\be
\langle 0| \frac{N_f}{16\pi^2} G_{\mu\nu}\tilde{G}_{\mu\nu}|\eta'\rangle = m_{\eta'}^2 f_{\pi}.
\ee
This does not mean that $\eta'$ is a glueball, even though the matrix element is larger than the matrix element of the isovector axial-vector
current for the pion
\be
\langle 0| \partial_{\mu} A_{\mu}^3|\pi\rangle = m_{\pi}^2 f_{\pi},
\ee
due to the larger $\eta'$ mass as compared to pion. In fact, the flavor-mixing angle between $\eta_1$ and $\eta_8$ for $\eta, \eta'$ have been well
studied with the help of axial anomaly~\cite{fks98}. Including the glueball mixing from the KLOE experiment of $\phi \rightarrow \gamma \eta,
\gamma \eta'$, the matrix elements of  $\langle 0| \frac{N_f}{16\pi^2} G_{\mu\nu}\tilde{G}_{\mu\nu}|M\rangle$
 for $M = \eta, \eta'$ and glueball $G$ are found
to be of the same order, even though in the large $N_c$ analysis, the matrix elements for
$\eta, \eta'$ are parametrically smaller
by $O(1/\sqrt{N_c})$ than that of the glueball~\cite{cll09}. This is known to be related to anomaly. On the other hand, the matrix element
$\langle 0| \bar{q} \gamma_5 q |G \rangle$ is more than an order of magnitude smaller than those of
$\langle 0| \bar{q} \gamma_5 q |\eta, \eta' \rangle (q = u, d, s)$~\cite{cll09}. This shows that the lower-dimension quark field
operators couple to the glueball much weaker than to the $q\bar{q}$ mesons. This has been taken as a criterion to distinguish the glueball
from the $q\bar{q}$ mesons under the condition that the glueball does not mix with the  $q\bar{q}$ mesons strongly.

In view of the above analysis
of the pseudoscalar mesons, it is suggested~\cite{liu08} that the smallness of the matrix element of lower-dimension quark operator compared to those of
established $q\bar{q}$ mesons is a better signal for the glueball than those with the higher dimensional glue operators. By the same token, we shall
adopt a similar criterion for detecting the hybrids by examining the dimension-4 $D$-type matrix element
$\langle 0|\bar{\psi} \Gamma \times \protect\nblr\psi|M\rangle$
and the dimension-5 B-type matrix element  $\langle 0|\bar{\psi} \Gamma \times B \psi|M\rangle$ of the
$1^{-+}$ and compare them with those
of the other ordinary mesons. If the $D$ matrix element of $1^{-+}$ is much smaller than others and the $B$ matrix element
much larger than (or at least as large as) the others, then it is a hybrid. Otherwise, it is not. Special attention will be paid to the comparison
with the $2^{++}$ meson. Neither $1^{-+}$ nor $2^{++}$ has dimension-3 interpolation field and their
dimension-4 operators are in the same Lorentz
multiplet i.e. $\bar{\psi} \gamma_{\mu} \nblr_{\nu} \psi$.

Using the vacuum-to-meson matrix element to discern the hybrid nature of the meson has been
adopted by Dudek~\cite{dud11} where a variational calculation with different dimensional operators
is carried out for mesons. It is asserted that overlap with the dim 5 B-type indicates hybrid-like
character~\cite{dud11}. This criterion, which was implicitly adopted by other lattice
calculations~\cite{lmb97,bhd97,jkm99,lm02,bbg03,ml03,hlw04,dep09}, faces several problems.
First of all, the transition matrix elements Z were compared only among the states
(ground and excited) of mesons with the same $J^{PC}$. For a variational calculation with a finite number
of operators, the matrix elements for one particularly operator will bound to have a largest value for
one of the states in the excitation spectrum. Therefore, there will always be a hybrid,
by definition, for each $J^{PC}$ which has any overlap with the $\bar{q}q D_{J=1}^{[2]}$ operator
(i.e. dim 5 B-type operator in our notation) involving the gauge field tensor G. This is hardly a test
to discern whether a state is a hybrid or not. Particularly, when these matrix elements are {\it normalized} in such a way that the largest value is set to unity for each of the operators used, there is no way to compare
Z from different operators for the same state, as they (having different derivatives with
d = 0,1,2,3) have different dimensions.  Instead, one should at least compare the matrix elements
of the $\bar{q}q D_{J=1}^{[2]}$ (dim 5) and $\bar{q}q D_{J=1}^{[1]}$ (dim 4) operators between
$1^{-+}$ and $2^{++}$. But this is not done. Secondly, as we stressed earlier, one cannot naively judge the nature of a state by the appearance of the interpolation field. We used the topological operator $G\tilde{G}$
as an example for illustration. According to many phenomenological and experimental analysis of the
matrix elements $<0|G\tilde{G}|\eta>, <0|G\tilde{G}|\eta'>$, and $<0|G\tilde{G}|glueball>$, it is
found that, in some solutions, $\eta$ and $\eta'$ matrix elements are larger than that of the
glueball~\cite{fks98,cll09}.
 This is not surprising as this is how U(1) anomaly is resolved in terms of the topological susceptibility in the Witten and Veneziano large $N_c$ approach. But according to the proposal in
Ref.~\cite{dud11} and, for that matter, many works on the subject, $\eta$ and $\eta'$ should be classified as glueballs, irrespective how strongly these state couple to the quark interpolation field with the dim 3
$\bar{q}\gamma_5 q$ operators. This serves as a counter example for this criterion.  Moreover, this criterion breaks down for pion as noted in Ref.~\cite{dud11}.  It is found~\cite{dud11} that the Z factors of the lowest pion state are the largest for both the $\bar{q}\gamma_5 q$ operator (dim 3) and the
$\rho_{NR} \times D_{J=1}^{[2]}$ (dim 5) operators. According to the proposed criterion~\cite{dud11}, the pion should be a hybrid. To avoid these difficulty and have a credible and practical criterion to distinguish a hybrid from the ordinary mesons, we think it is essential to compare matrix elements for the operators of
the same dimension across the board of different mesons. This is what we propose to do.

\subsection{Non-relativistic Operators}\label{subsec:non-rela}
To address the question of the exotic quantum number, it would be useful to find out the non-relativistic form of the interpolation operators listed
in Sec.~\ref{sec:interpolation}. We use Foldy-Wouthuysen-Tani transformation~\cite{fwt50} for non-relativistic reduction to the
heavy quark and anti-quark fields described by the Pauli spinors $\phi$ and $\chi$.

 \begin{table}[hbt]
 \caption{Non-relativistic form for the three kinds of operators ($\Gamma, D$ and $B$) as shown in Table 1.
Here we list the operators $\mathcal{O}$ in  the interpolation field $\chi^{\dagger}\mathcal{O}\phi$. Repeated indices are summed over.
 }\label{tab:non-Oper}
 \begin{center}
 \begin{tabular}{c|c|c|c}
\hline
            &$\Gamma$   & $D$      & $B$ \\
\hline
$0^{-+}$  &$\mathbb{I}$                         &$\frac{1}{2m_c}\nblr_i\nblr_i$           &$i\sigma_iB_i$ \\
$1^{--}$  &$\sigma_i$                           &$\frac{1}{2m_c}\sigma_j\nblr_j\nblr_i$   &$B_i$   \\
$0^{++}$  &$\frac{1}{2m_c}\nblr_i\sigma_i$        &$\sigma_i \nblr_i$         &$\frac{1}{2m_c}\nblr_iB_i$ \\
$1^{++}$  &$\frac{1}{2m_c}\eps\nblr_j\sigma_k$    &$\eps\sigma_j\nblr_k$      &$\frac{1}{2m_c}(\eps\nblr_jB_k+\emph{i}\partial_i(\sigma_jB_j))$\\
$1^{+-}$  &$\frac{1}{2m_c}\nblr_i$                &$\nblr_i$                  &$\frac{1}{2m_c}\sigma_j\nblr_j B_i$          \\
$2^{++}$  &                                     &$|\eps|\sigma_j\nblr_k$    &$\frac{1}{2m_c}|\eps|(\nblr_jB_k +\emph{i}\varepsilon_{jmn}\sigma_m\partial_n(B_k))$\\
$1^{-+}$  &                                     &$\frac{1}{2m_c}(\sigma\cdot\nbl\nblr_i+\nblr_i\sigma\cdot\nbr)$  & $\eps\sigma_jB_k$\\
          &                                     &$\frac{1}{2m_c}(\nbl_i\sigma_j\nblr_j+\sigma_j\nblr_j\nbr_i)$ &\\
\hline
 \end{tabular}
 \end{center}
 \end{table}

The Dirac spinor $\psi$ and $\bar{\psi}$ are expanded
in terms of $\phi$ and $\chi$ in $1/m$ as
 \ba
 &\psi=&e^{\frac{\gamma\cdot D}{2m}}\Big(\begin{array}{c} \phi \\ \chi \end{array}\Big)
 =\Big[1+\frac{\gamma\cdot D}{2m}+\frac{\gamma\cdot \vec{D}\ \gamma\cdot D}{8m^2}O(1/m^3)\Big]\Big(\begin{array}{c} \phi \\ \chi \end{array}\Big)\nonumber\\
 &\ &=\Big(\begin{array}{c} \phi \\ \chi \end{array}\Big)+
 \frac{\emph{i}}{2m}\Big(\begin{array}{c} -\sigma\cdot\vec{D}\chi \\ \sigma\cdot\vec{D}\psi \end{array}\Big)+\frac{(\nbr^2)}{8m^2}\Big(\begin{array}{c}
\phi \\ \chi \end{array}\Big)+O(1/m^3),\\
  &\bar{\psi}=&\Big(\begin{array}{cc} \phi^{\dagger} & -\chi^{\dagger} \end{array}\Big)e^{-\frac{\gamma\cdot \overleftarrow{D}}{2m}}=\Big(\begin{array}{cc}
\phi^{\dagger} & -\chi^{\dagger} \end{array}\Big)+
 \frac{\emph{i}}{2m}\Big(\begin{array}{cc} \chi^{\dagger}\sigma\cdot\overleftarrow{D}^{\dagger} & \phi^{\dagger}\sigma\cdot\overleftarrow{D}^{\dagger} \end{array}\Big)\\
 &&+\frac{(\nbl^2)}{8m^2}\Big(\begin{array}{cc} \phi^{\dagger} & -\chi^{\dagger} \end{array}\Big)+O(1/m^3),\\
 &\textrm{where} &\gamma_i=\Big(\begin{array}{cc}0&-\emph{i}\sigma_i\\\emph{i}\sigma_i&0\end{array}\Big),\
 \gamma_4=\Big(\begin{array}{cc}I&0\\0&-I\end{array}\Big),\ \gamma_5=\Big(\begin{array}{cc}0&I\\I&0\end{array}\Big),\nonumber\\
 &&\Sigma_i\equiv\varepsilon_{ijk}\sigma_{jk}=\Big(\begin{array}{cc}\sigma_i&0\\0&\sigma_i\end{array}\Big),\
 D_i=\partial_i+\emph{i}A_i^aT^a
 \ea

Operator $D$ used here is the spatial part of the covariant derivative and $m$ is the heavy-quark mass. The Pauli spinors $\phi/\phi^{\dagger}$ and
$\chi^{\dagger}/\chi$ are the annihilation/creation operators for the heavy quark and antiquark which satisfy the relation
 \be
 \phi|0\rangle = \chi^{\dagger}|0\rangle = 0; \hspace{0.5cm} \langle 0|\phi^{\dagger}= \langle 0|\chi=0.
 \ee

With the above approximation, we could reduce the operators listed in Table~\ref{tab:Operators} with a given $J^{PC}$ to the form of
$\chi^{\dagger}\mathcal{O}\phi$ and $\phi^{\dagger}\mathcal{O}^{\dagger}\chi$ with $\mathcal{O}$ now involves $\sigma, \nblr,$ and $B$. We shall
still classify them according to their dimensions and label them the same as before, i.e.
$\Gamma$-type (dimension 3), $D$-type (dimension 4), and $B$-type (dimension 5). The operators for $\chi^{\dagger}\mathcal{O}\phi$ to leading order in $1/m$ are
listed in Table\ \ref{tab:non-Oper}. Note that $\nblr$ acts on the quark and anti-quark fields, while $\partial$ acts on the glue field $B$.

\subsection{Exotic quantum numbers}

       We see from the non-relativistic reduction in the above section that the dimension-4 \mbox{($D$-type)} interpolation field for
the $1^{-+}$ meson involves a symmetric combination of $\nbl$ and $\nbr$. This is the center of mass momentum operator of the $q\bar{q}$ pair.
We now see why this operator is not admissible in the quark model with only the constituent quark degree of freedom.
In this model, the center of mass of  $q\bar{q}$
is not a dynamical variable due to translational invariance, while the quantum number $J^{PC}$ is defined in the center of mass of the  $q\bar{q}$ pair.
In QCD, on the other hand, there are gluons besides the quarks so that the $q\bar{q}$ pair can have orbital angular momentum relative to the glue stuff, much like the
orbital motion of the electron pairs around the nucleus in the atom, or the planetary motion of the earth-moon pair around the sun. This is also true in models
where there are other constituents that the $q\bar{q}$ pair can recoil against. For example, in the MIT bag model, the $q\bar{q}$ can have orbital
angular momentum against the bag if the latter is made dynamical~\cite{reb75}. In the chiral quark model, the  $q\bar{q}$ can recoil against the pion. In the context of the flux-tube model which is a good and appropriate picture for
heavy quarkoniums, the P-wave quarkonium is pictured to have the flux-tube rotate in phase with the heavy quark and
antiquark at its opposite ends. Since the flux-tube is not excited internally with transverse vibration, it is not a
hybrid in the flux-tube model~\cite{ip85}. By the same token, one can picture the heavy $1^{+}$ meson with the
flux-tube folding up so that the the center of mass of the $q\bar{q}$ pair rotates against the folded flux-tube with no vibrational excitation of the tube.

       In fact, the issue of the the exotic quantum number and its relation to the center-of-mass motion of the $q\bar{q}$ has been raised in
the MIT bag model~\cite{jj76,dj76}. An example is given for the $2^{+\pm}$ meson where the quark and anti-quark orbital wavefunctions are given as
\be
\Psi(2^{+\pm}) = \frac{1}{\sqrt{2}} (S_{\frac{1}{2}}\overline{P}_{\frac{3}{2}} \mp P_{\frac{3}{2}}\overline{S}_{\frac{1}{2}}).
\ee
Since both $C = \pm$ are possible, they double the spectrum from the conventional constituent $q\bar{q}$ model. It is pointed out that the symmetric
combination leads to a $P$-wave for the center of mass of the $q\bar{q}$. In the nuclear shell model with harmonic oscillator potential,
this is considered a spurious center-of-mass excitation since the center of mass is pinned down by the harmonic oscillator potential. If the
bag is not dynamical like the external harmonic oscillator in the shell model, it can be removed with center-of-mass correction~\cite{lw82}.
However, if the bag is considered dynamical with surface fluctuations~\cite{reb75}, this center-of-mass motion is physical and so
is the $2^{+-}$ state. By analogy, one can consider the $1^{-\pm}$ states with the combination
\be
\Psi(1^{-\pm}) =  \frac{1}{\sqrt{2}} (1S_{\frac{1}{2}}2\overline{S}_{\frac{1}{2}} \mp 2S_{\frac{1}{2}}1\overline{S}_{\frac{1}{2}}),
\ee
with the anti-symmetric combination being the $1^{-+}$ state where both the center-of-mass and relative coordinates are in the $P$-wave for harmonic oscillator wavefunctions.

      To conclude this part of the discussion, we see that the `exotic' quantum numbers exist in QCD and models with additional constituents besides
the  $q\bar{q}$ pair. The `exoticness' is only in the context of the constituent quark model with only  $q\bar{q}$ degree of freedom. These quantum
numbers can be accommodated with parity and total angular momentum from Eqs.~(\ref{PC}) and (\ref{J}) supplanted by
\bea   \label{PJ}
P & =& (-)^{L+l+1}   \nonumber \\
\vec{J} &=& \vec{L}+\vec{l} + \vec{S}.
\eea
where $L$ is the orbital angular momentum of the $q\bar{q}$ pair in the hadron. The charge parity \mbox{$C = (-)^{l +S}$} remains the same, provided that other degrees of freedom in the hadron are not excited and gives $C = +$. In the case of
$1^{-+}$, $L = l = S = 1$ and the two operators in Table~\ref{tab:non-Oper} correspond to  $\vec{S} + \vec{L} =0$ and
$\vec{S} + \vec{l} =0$ respectively. Other `exotic' quantum numbers, e.g. $0^{+-}, 2^{+-}, 3^{-+}$ can all be accommodated in Eq.~(\ref{PJ}).

\section{Numerical Details}

       We shall give lattice details including the action, the parameters as well as the operators used for the
interpolation fields of various mesons.

\subsection{Improved Clover Action}\label{subsec:action}

    We adopt the anisotropic Wilson gauge action \cite{Klassen:1998ua} in the quenched approximation for the present study.
The improved anisotropic Wilson fermion action is
\ba
 \label{eq:shift_M}
 M_{xy} &=&\delta_{xy}\sigma + {\mathcal A}_{xy}
 \nonumber \\
 {\mathcal A}_{xy} &=&\delta_{xy}\left[1/(2\kappa_{max})
 +\rho_t \sum^3_{i=1} \sigma_{0i} {\mathcal F}_{0i}
 +\rho_s (\sigma_{12}{\mathcal F}_{12} +\sigma_{23}{\mathcal F}_{23}
 +\sigma_{31}{\mathcal F}_{31})\right]
 \nonumber \\
 &-&\sum_{\mu} \eta_{\mu} \left[
 (1-\gamma_\mu) U_\mu(x) \delta_{x+\mu,y}
 +(1+\gamma_\mu) U^\dagger_\mu(x-\mu) \delta_{x-\mu,y}\right] \;\;,
 \ea
 where the coefficients are given by
 \ba
 \label{eq:parameters}
 \eta_i &=&{\nu \over 2u_s} \;\;, \;\;
 \eta_0={\xi\over 2} \;\;,
 \;\;\sigma={1 \over 2\kappa}-{1\over 2\kappa_{max}}\;\;,
 \nonumber \\
 \rho_t &=& \nu{(1+\xi)\over 4u^2_s} \;\;, \;\;
 \rho_s = {\nu \over 2u^4_s} \;\;.
 \ea
with $\xi=a_s/a_t$ being the bare aspect ratio of the asymmetric lattice, and $\nu$ the bare speed of light parameter.
Another parameter $u_s$, taken to be the fourth root of the average spatial plaquette value, is used to incorporate the tadpole improvement
of the spatial gauge link $U_i(x)$.

With this fermion action, the bare mass of the quark is
 \be
 \label{eq:kcr}
 m_0a_s={1\over 2\kappa}-\xi-3\nu\;.
 \ee

The lattice used in this study is of the size $12\times12\times12\times96$ at $\beta=2.8$ which gives $a_s=0.138$ fm,
with the aspect ratio $\xi=a_s/a_t=5$.

The bare $\kappa$ of the charm quark is set to 0.060325 with the bare speed of light parameter $\nu=0.74$, which is determined by fitting the mass of $J/\psi$.
Similarly, The bare $\kappa$ of the strange quark is set to 0.0615 which gives the vector mass close to that of $\phi$.

\subsection{Masses and Vacuum to Meson Transition Matrix Elements} \label{subsec:amp}

\begin{figure}
{\includegraphics[scale=1.0]{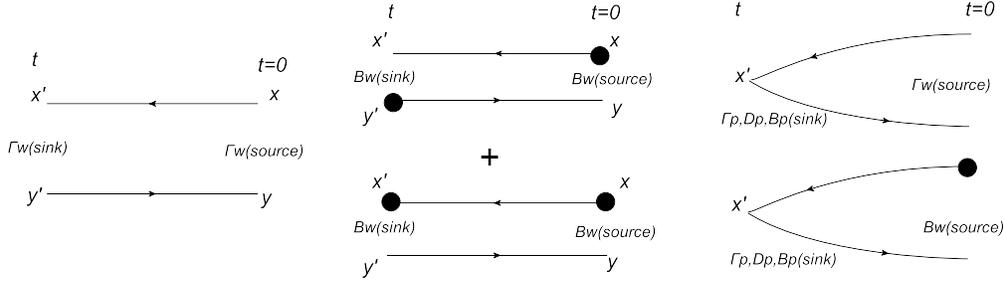}}
\caption{Sketch of two point functions. The lines denote quark and antiquark propagators. The black dot is the glue field tensor $B$ attached
at the quark wall source and sink.}\label{fig:diagram}
\end{figure}

To construct two-point functions, we use the $\Gamma$-type wall operators for mesons which have dimension-3 interpolation fields. For those which
do not have dimension-3 interpolation fields, we use the B-type wall source to enhance the signals. This is illustrated in Fig.\ \ref{fig:diagram}.
$B_w$ denotes the wall source and sink for the $B$-type operator. We note that the glue field tensor $B$ can be attached to either the
quark field or the antiquark field at the source and sink. It is indicated by a black dot in the figure.
When both the wall source and sink are of the $B$-type operators, it is necessary to sum the two kinds of diagrams (middle one in
Fig.\ \ref{fig:diagram})
to obtain an eigenstate of charge parity. When the sink is the point operator which has a definite charge-parity, one diagram with the
$B$ attached to either the quark or anti-quark wall source (the right diagram in Fig..\ \ref{fig:diagram}) will suffice.
$\Gamma_p$, $D_p$ and $B_p$ denote the point sinks.


The wall-source is placed on 16 of the total 96 time slices separately for each of the 1000 configurations to gain statistics.
We calculate the correlators with both the source and sink being the wall $B$-type operators ($B_w$) and with $B$-wall and point sinks with the $\Gamma, D$ and $B$ operators ($\Gamma_p, D_p$ and $B_p$). The color magnetic field $B$ is smeared twice
for the wall source and sink and the double antisymmetric derivative operator
$\eps \bar{\psi}\nblr_j\nblr_k\psi$ is used to replace $\bar{\psi}\psi B_i$ for the point sink.

The ground state mass and the vacuum to meson transition matrix element are extracted from the following correlators:

 \ba\label{eq:amp}
 && \langle [\bar{\psi}(\Gamma \times B) \psi]_{wall }^{\dagger}(t)\,[\bar{\psi}(\Gamma \times B) \psi]_{wall }(0)\rangle_{\stackrel{---\longrightarrow}
{t \gg a}}  N_V \frac{|\langle 0|[\bar{\psi}(\Gamma \times B) \psi]_{wall}|J^{PC}\rangle|^2}{2m}\, (e^{-m t}+e^{-m(n_T-t)}),\nonumber\\
 && \langle \mathcal{O}_p(t)\, [\bar{\psi}(\Gamma \times B) \psi]_{wall}(0)\rangle_{\stackrel{---\longrightarrow}{t \gg a}}
 N_V \frac{\langle 0| \mathcal{O}_p|J^{PC}\rangle \langle J^{PC}|[\bar{\psi}(\Gamma \times B) \psi]_{wall}|0\rangle}{2m}\,
(e^{-m t}+e^{-m(n_T-t)}),\nonumber\\
\ea
where $N_V = L^3$ is the three-volume factor. From these two equations, one can obtain the matrix element
$\langle 0|\mathcal{O}_p|J^{PC}\rangle$.
\be
\langle 0|\mathcal{O}_p|J^{PC} \rangle =\lbrack \frac{2m (e^{-m t}+e^{-m(n_T-t)})}
{N_V \langle [\bar{\psi}(\Gamma \times B) \psi)]_{wall }^{\dagger}\, [\bar{\psi}(\Gamma \times B) \psi]_{wall }\rangle}\rbrack^{1/2}
\langle  \mathcal{O}_p\, [\bar{\psi}(\Gamma \times B) \psi]_{wall}\rangle.
\ee
 Similarly, we also obtain the masses and $\Gamma$-type matrix elements with the $\Gamma$-type wall source
(right and left diagrams in Fig.~\ref{fig:diagram}).

\section{Numerical results and Discussion}\label{sec:res}

\subsection{Charmoniums}

   We first calculate the masses and the matrix elements for the charmonium with the
charm quark $\kappa = 0.060325$
that was tuned to the physical $J/\Psi$ mass. The masses from different correlators are listed in
Table~\ref{tab:cmass}.

 \begin{table}[hbt]
\caption{Masses of charmonium states from $\Gamma$- and $B$-type sources and point sinks.}
 \label{tab:cmass}
 \begin{center}
 \begin{tabular}{c|ccccc}
\hline
    &$\Gamma_w\to \Gamma_p$&$ B_w\to \Gamma_p $&$ B_w\to B_p $&$ B_w\to D_p$&$\textrm{PDG}$\\
\hline
$0^{-+}$&$3000\pm\ 3 $&$3000\pm\ 3   $&$2999\pm\ 3   $&$3000\pm\ 3   $&$2980.3\pm1.2$\\
$1^{--}$&$3096\pm\ 3 $&$3095\pm\ 3   $&$3093\pm\ 3   $&$3094\pm\ 3   $&$3096.916\pm0.011$\\
$0^{++}$&$3458\pm30  $&$3485\pm18    $&$3485\pm21    $&$3476\pm18    $&$3414.75\pm0.31$\\
$1^{++}$&$3497\pm21  $&$3491\pm10    $&$3492\pm28    $&$3492\pm28    $&$3510.66\pm0.07$\\
$1^{+-}$&$3489\pm30  $&$3475\pm21    $&$3486\pm12    $&$3494\pm\ 6   $&$3525.42\pm0.29$\\
$2^{++}$&--         &--           &$3529\pm40    $&$3501\pm13    $&$3556.20\pm0.09$\\
$1^{-+}$&--         &--           &$4205\pm84    $&$4234\pm42    $& --\\
\hline
\end{tabular}
 \end{center}
 \end{table}

\begin{figure}[h]
\begin{minipage}{0.5\linewidth}
{\includegraphics[scale=0.6]{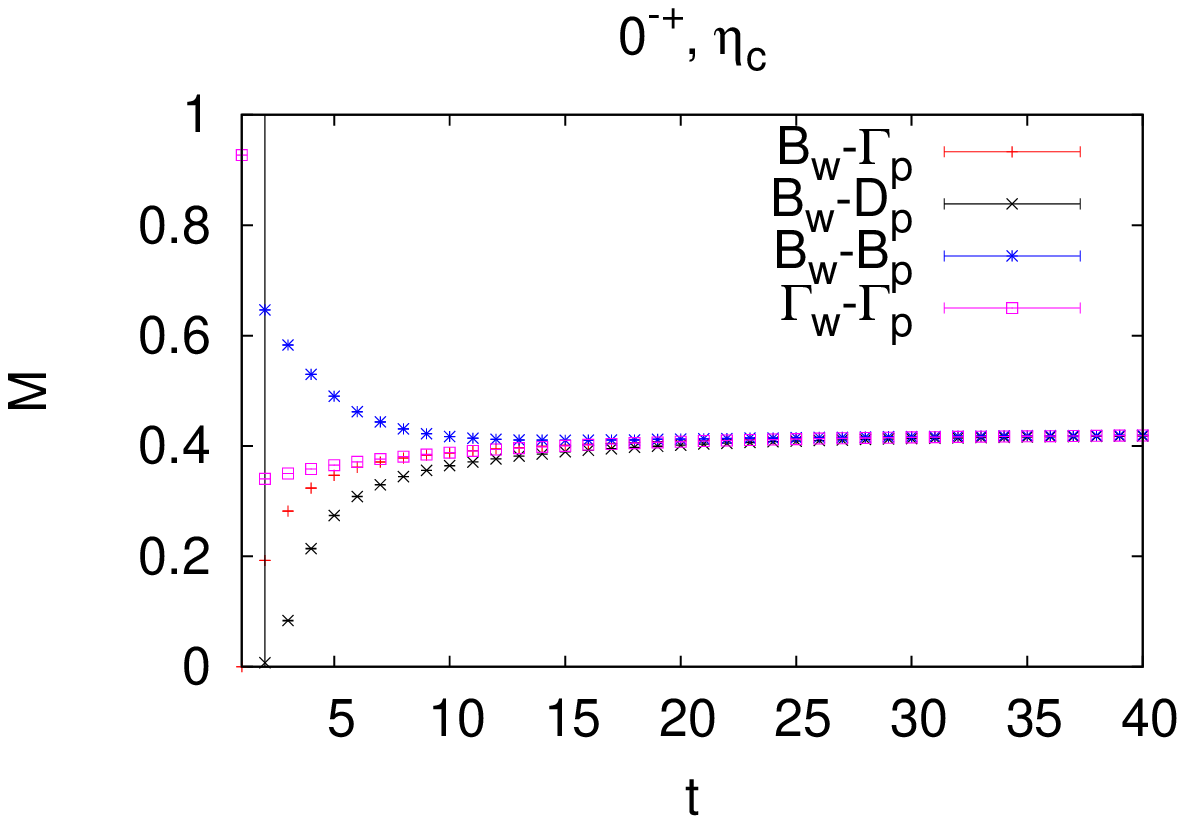}}
\end{minipage}%
\begin{minipage}{0.5\linewidth}
{\includegraphics[scale=0.6]{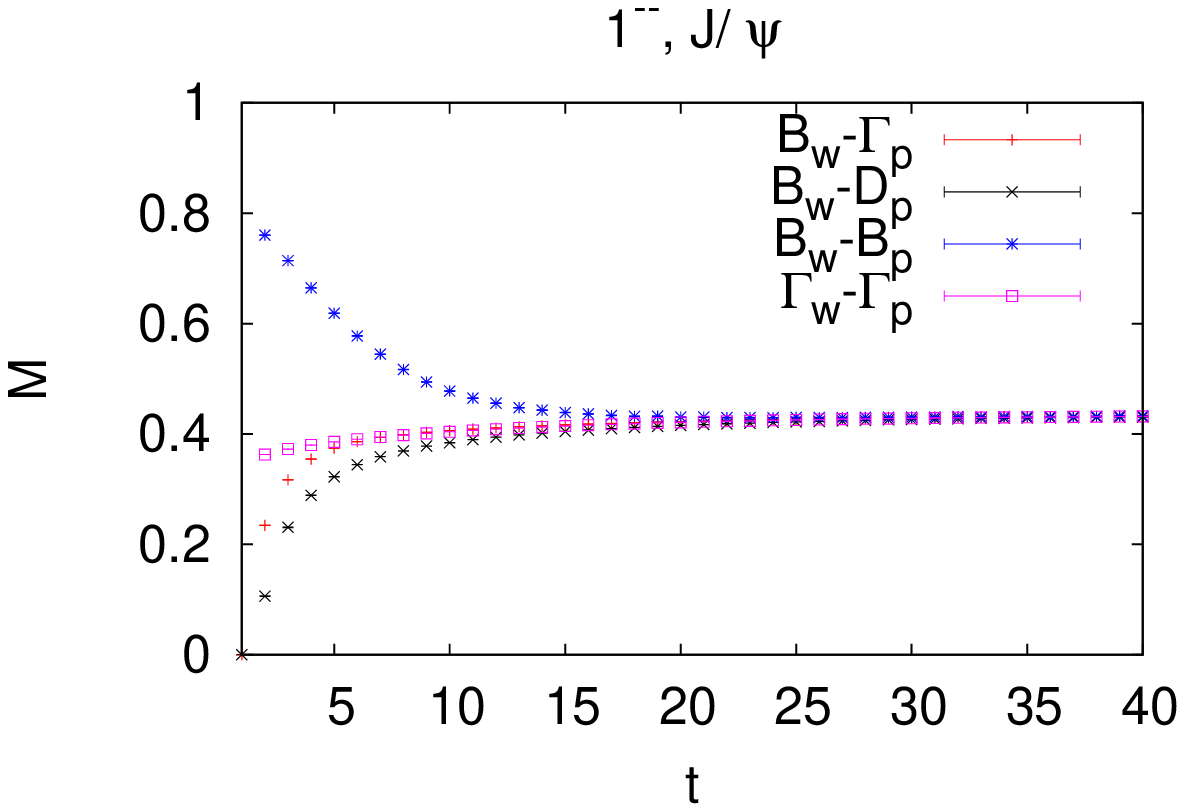}}
\end{minipage}
\caption{\small Effective mass plot for $\eta_c$ and $J/\Psi$ with $B$- and $\Gamma$-type operators as the  wall sources and $\Gamma_p, D_p$ and $B_p$ for the zero momentum point sinks.}\label{fig:swave}
\end{figure}

\begin{figure}[h]
\begin{minipage}{0.5\linewidth}
{\includegraphics[scale=0.6]{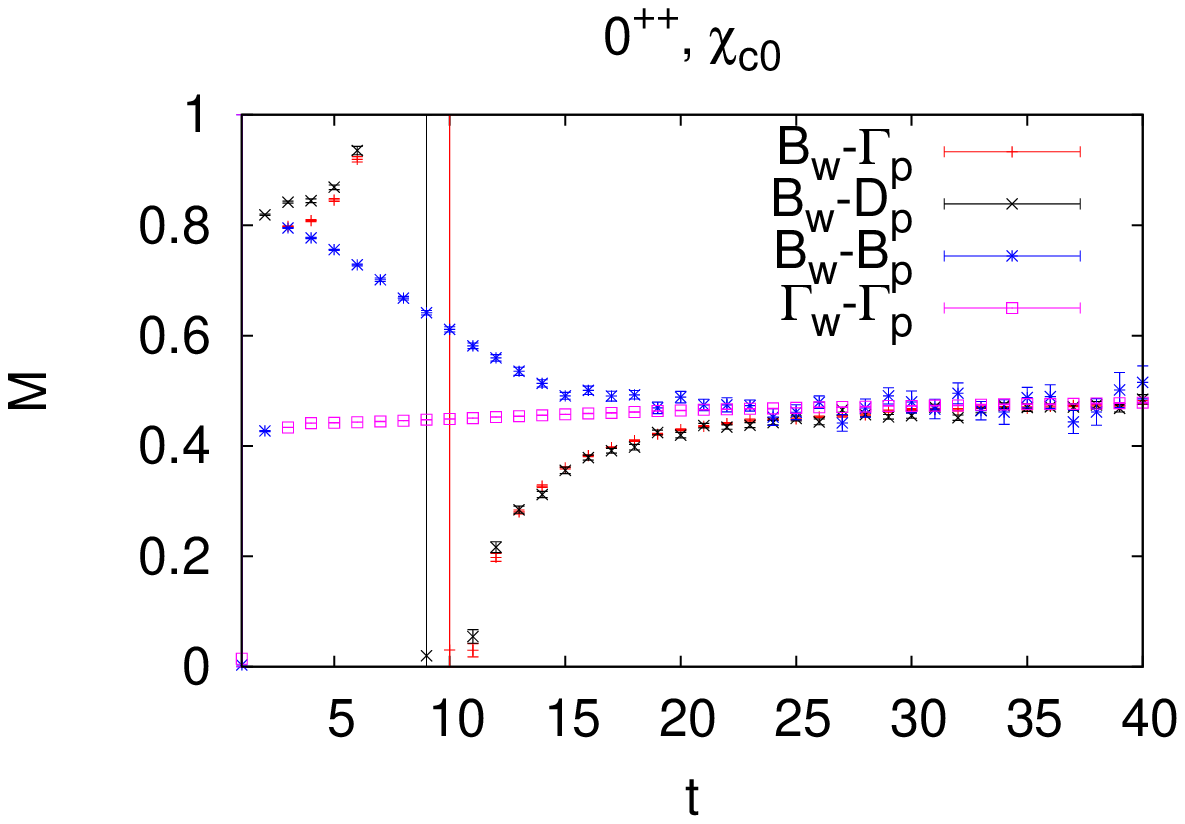}}
\end{minipage}%
\begin{minipage}{0.5\linewidth}
{\includegraphics[scale=0.6]{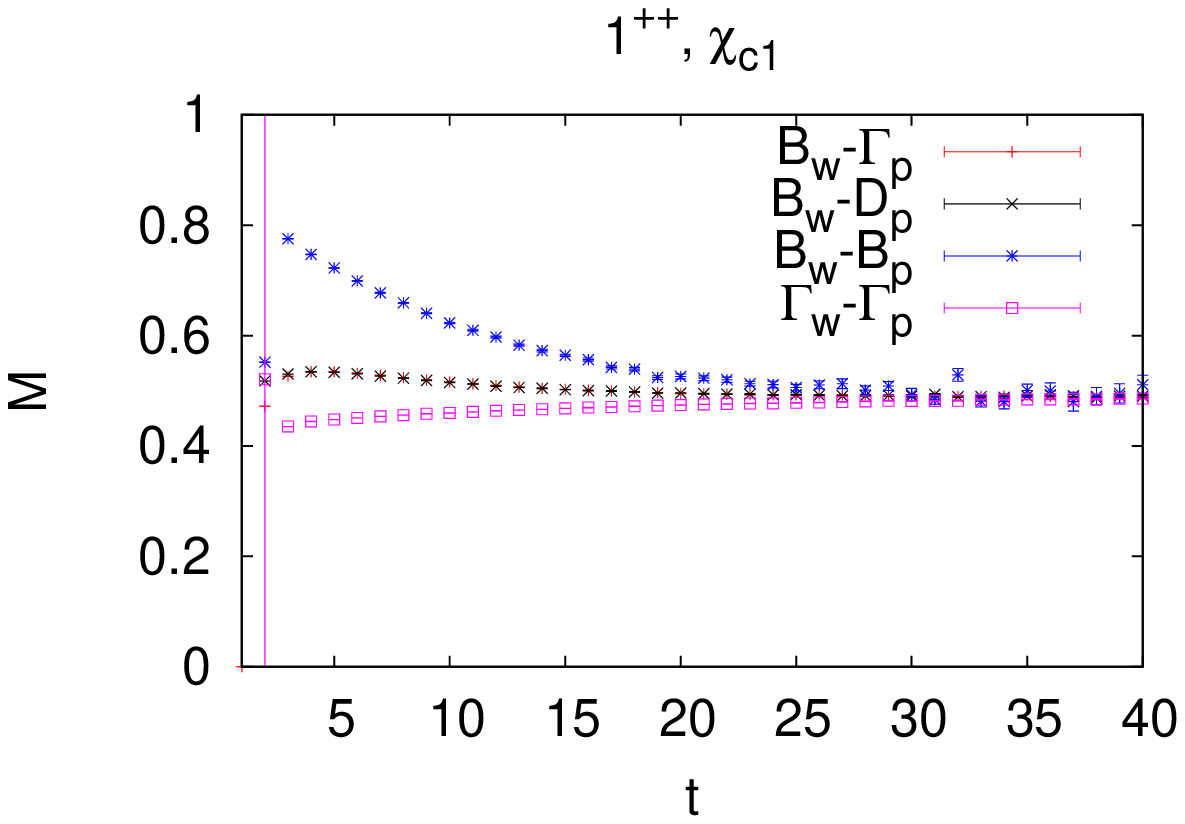}}
\end{minipage}
\caption{\small The same as Fig.~\protect\ref{fig:swave} for $\chi_{c0}$ and $\chi_{c1}$}\label{fig:pwave1}
\end{figure}

\begin{figure}[h]
\begin{minipage}{0.5\linewidth}
{\includegraphics[scale=0.6]{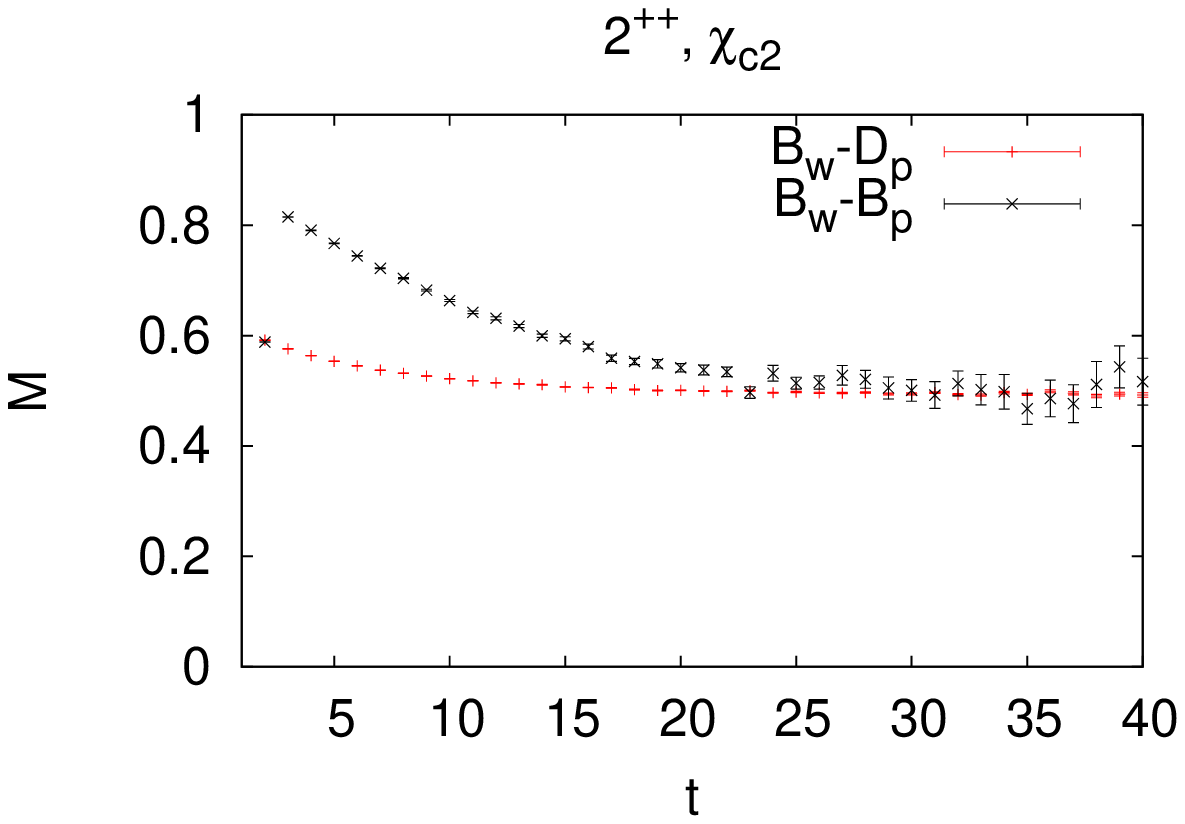}}
\end{minipage}%
\begin{minipage}{0.5\linewidth}
{\includegraphics[scale=0.6]{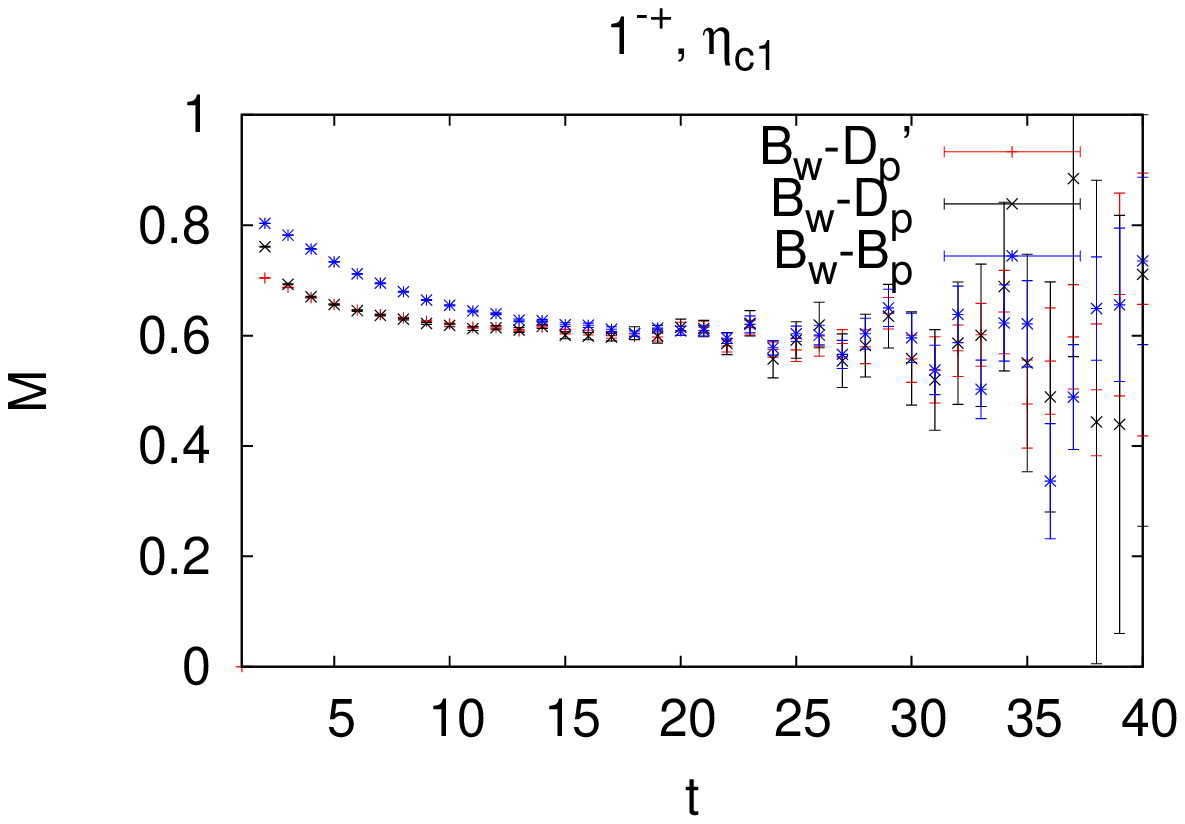}}
\end{minipage}%
\caption{\small The same as Fig.~\protect\ref{fig:swave} for $\chi_{c2}(2^{++})$ and
$\eta_{c1} (1^{-+})$} \label{fig:pwave2}
\end{figure}

The effective masses of $\eta_c (0^{-+})$ and $J/\Psi (1^{--})$ are plotted in Fig.~\ref{fig:swave} with the
$B$- and $\Gamma$-type wall sources and $\Gamma_p, D_p$ and $B_p$ for the zero momentum point sinks.
The effective masses of $\chi_{c0} (0^{++})$ and $\chi_{c1} (1^{++})$ from the wall sources are plotted in
Fig.~\ref{fig:pwave1}
for several point sinks. The effective masses of $\chi_{c2} (2^{++})$ and $\eta_{c1} (1^{-+})$ from the wall sources are plotted in Fig.~\ref{fig:pwave2} for several point sinks.

We see that the masses obtained from different correlators with different sources and sinks are all consistent with each other and
the pattern of the charmonium masses, besides $1^{-+}$, are in reasonable agreement with experiments, except the hyperfine
splitting which is known to be smaller than experiment for the quenched approximation~\cite{tac06}. We note that
dimension-4 and -5 operators produce the same mass of $1^{-+}$ within errors. We take this to imply that they are the same state.

The effective masses of the pseudoscalar ($\eta_c$) and vector ($J/\Psi$) charmonium are plotted in Fig.~\ref{fig:swave} for
the cases with $\Gamma_w$ and $B_w$ sources and $\Gamma_p, D_p$ and $B_p$ sinks.
The effective masses for the scalar ($\chi_{c0}$) and axial-vector ($\chi_{c1}$) are plotted in Fig.~\ref{fig:pwave1} and
those for the tensor ($\chi_{c2}$) and $1^{-+}$ are plotted in Fig.~\ref{fig:pwave2}. As we can see from Table~\ref{tab:cmass},
they agree for different sources and sinks within errors. Different interpolation fields project to the same lowest states
in all channels studied here.

Before we discuss the results on the matrix elements, we should point out a relation between the
dimension-4 $D$-type and dimension-5 $B$-type operators in the non-relativistic limit.

The double derivative operator in the leading non-relativistic expansion of the $1^{-+}$ interpolation field can
be expanded as
\bea\label{eq:dd_reduce}
&&\chi^{\dagger}(\nbl_i\nblr_j+\nblr_j\nbr_i)\psi\nonumber\\
&=&\chi^{\dagger}(\nl_i.\nl_j-\nr_i.\nr_j-\nl_i.\nr_j+\nl_j.\nr_i)\psi\nonumber\\
&&+\chi^{\dagger}(2\emph{i}.\partial_j(A_i)-2[A_i,A_j])\psi+\chi^{\dagger}(2\emph{i}(\nl_i.A_j+\emph{i}.A_j.\nr_i)\psi
\eea
Since we are projecting to the zero momentum meson state in the lattice calculation with periodic condition in the
spatial direction, we have
\bea \label{eq:tot_der}
\int\! d^3x \,\chi^{\dagger}\nl_i.\nl_j\psi= \int\! d^3x \,\chi^{\dagger}\nr_j.\nr_i\psi,
\ \int\! d^3x \,\chi^{\dagger}\nl_i.\nr_j\psi= \int\! d^3x \,\chi^{\dagger}\nl_j.\nr_i\psi,\nonumber\\
\int\! d^3x \, \partial_i(\chi^{\dagger}A_j\psi)=\int\! d^3x \,\chi^{\dagger}(\nl_i.A_j+A_j.\nr_i)\psi+
\chi^{\dagger}\partial_j(A_i)\psi = 0.
\eea
From Eqs.\ (\ref{eq:dd_reduce}) and (\ref{eq:tot_der}), we obtain
\bea
&&\int\! d^3x \, \chi^{\dagger}(\nbl_i\nblr_j+\nblr_j\nbr_i)\psi= \int\! d^3x\,
\chi^{\dagger}(2\emph{i}.\partial_j(A_i)-2\emph{i}.\partial_i(A_j)-2[A_i,A_j])\psi\nonumber\\
&=& \int\! d^3x \, 2\emph{i}\chi^{\dagger}G_{ij}\psi
\eea

Therefore the leading non-relativistic terms of the zero-momentum $1^{-+}$ interpolation fields are
 \ba   \label{eq_1mp_non_relat}
\int\! d^3x\, \bar{\psi}^a\gamma_4\nblr_i \psi^a &{}_{\stackrel{---\longrightarrow}{N.R.}}&
 \frac{\emph{i}}{2m}\int d^3x\,\chi^{\dagger}(\sigma\cdot\nbl\nblr_i+\nblr_i\sigma\cdot\nbr)\phi\nonumber\\
&=  & \frac{1}{2m}\int d^3x\,\chi^{\dagger}2\sigma_j G_{ji}\phi
=-\frac{1}{2m}2\varepsilon_{ijk}\int d^3x\,\chi^{\dagger}\sigma_jB_k\phi\\
\int\! d^3x\, \bar{\psi}^a\varepsilon_{ijk}\Sigma_j\nblr_k \psi^a &{}_{\stackrel{---\longrightarrow}{N.R.}}&
 \frac{\emph{i}}{2m}\int d^3x\, \chi^{\dagger}(\nbl_i\sigma\cdot\nblr+\sigma\cdot\nblr\nbr_i)\phi\nonumber\\
&=&\frac{1}{2m}\int d^3x\, \chi^{\dagger}2\sigma_j G_{ij}\phi
=\frac{1}{2m} 2\varepsilon_{ijk} \int d^3x\,\chi^{\dagger}\sigma_jB_k\phi.
 \ea

We see that, up to a sign and a proportional constant (i.e. heavy quark mass $m$), both dimension-4 $D$-type operators of $1^{-+}$ are
equivalent to the dimension-5 $B$-type operator with the magnetic field in the non-relativistic limit. The matrix elements
from all three operators are expected to be the same up to a known constant and $\mathcal{O}(a)$ for heavy quarkoniums.

The matrix elements for the charmoniums are listed in Table~\ref{tab:amp_charm}.

\begin{table}[htb]
 \caption{Matrix elements $<0|\mathcal{O}_p|J^{PC}>$ for charmoniums.
 }\label{tab:amp_charm}
 \begin{center}
 \begin{tabular}{c|l|l|l}
\hline
    &$\Gamma_p$ & $D_p$ & $B_p$\\
\hline
$0^{-+}$  &$0.0697\pm0.0014$    &$0.0503\pm0.0007$   &$0.0251\pm0.0006$ \\
$1^{--}$  &$0.0502\pm0.0005$    &$0.0149\pm0.0001$   &$0.0075\pm0.0002$ \\
$0^{++}$  &$0.035\ \pm0.005$    &$0.075\ \pm0.015$  &$0.009\ \ \pm0.003$ \\
$1^{++}$  &$0.020\ \pm0.003$    &$0.062\ \pm0.005$  &$0.0023\pm0.0002$  \\
$1^{+-}$  &$0.014\ \pm0.002$    &$0.045\ \pm0.005$  &$0.0019\pm0.0002$  \\
$2^{++}$  &                     &$0.044\ \pm0.003$  &$0.00080\pm0.00008$ \\
$1^{-+}$  &                     &$0.0059\pm0.0005$  &$0.0082\pm0.0006$  \\
          &                     &$0.0054\pm0.0004$ & \\
\hline
\end{tabular}
 \end{center}
 \end{table}

The matrix elements of 0.0059(5) and 0.0054(4) for the two $D$ operators of $1^{-+}$ are the same which are expected from the above
discussion. Eq.~(\ref{eq_1mp_non_relat}) also shows that they should be $1/ma$ times that of the dimension-5 $B$-type to $\mathcal{O}(a)$.
On the anisotropic lattice used here, dimension-less $\frac{1}{ma}$ should be replaced by the anisotropic form,
\ba
\frac{1}{m a}\Rightarrow \frac{\nu}{m_c a_t\xi}\sim 0.7048
\ea
where $\xi$ and $\nu$ are defined in Eq.\ (\ref{eq:parameters}). Multiplying this factor to the $B$-type matrix element  0.0082(6) gives 0.0058(4) which agrees with the $D$-type matrix elements quite well.

\begin{figure}[ht]
\includegraphics[scale=1.0]{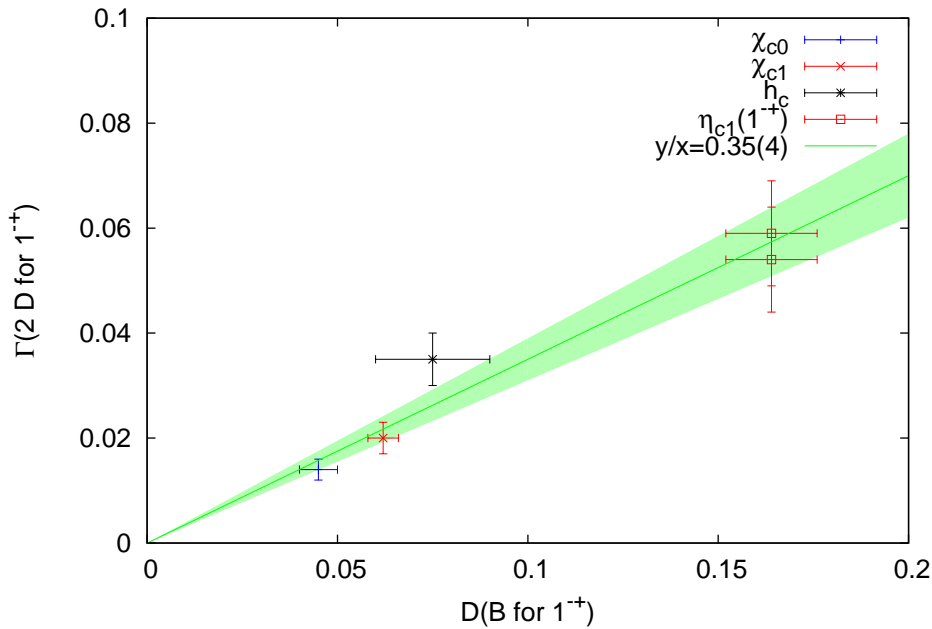}
\caption{\small Global fit for the ratios of $\Gamma$ for $\chi_{c0}, \chi_{c1}$ and $h_c$ (2 $D$ for $1^{-+}$) m.e.
to the corresponding m.e.of  $D$ ($B$ for  $1^{-+}$).}\label{fig:fit}
\end{figure}

Furthermore, comparing $\Gamma$ and $D$ operators for the P-wave states
$0^{++}, 1^{++}$ and $1^{+-}$ in Table~\ref{tab:non-Oper} shows that they are related by $\frac{1}{2m}$. Thus, we expect dimension-3 $\Gamma$ matrix elements to be
$\epsilon = \frac{1}{2m} = 0.3524 $ times the dimension-4 $D$ matrix elements. To check this, we plot
$2$ times the $\Gamma$ m.e. against
the $D$ matrix elements in Fig.~\ref{fig:fit} for these states and also $2 $ times the $D$
matrix elements of $1^{-+}$ meson against the
corresponding $B$ matrix elements. We fit the ratio of all the data and find the slope to be 0.35(4). This is quite consistent with
$\epsilon = 0.3524$. This shows that the matrix elements we studied for the charmonium states are quite non-relativistic in the
sense that higher orders in $1/m$ are not important to spoil the equivalence relation we found in Eq.~(\ref{eq_1mp_non_relat}) and that cutoff effect in $\mathcal{O}(a)$ is small. Since we are considering matrix
elements of operators with different dimensions, there is a concern about operator mixing. The results
in Fig.~\ref{fig:fit} suggest that the mixing effects between the
dim-3 $\Gamma$-type and the dim-4 $D$-type operators and also between the dim-4 $D$-type and
dim-5 $B$-type opearators are also small.

At first sight, the $D$ matrix elements of $1^{-+}$ are about an order of magnitude smaller than those of the other mesons. However, upon
comparing with $2^{++}$ in Table~\ref{tab:non-Oper}, we see that the $1^{-+}$ operators have an extra factor
of $(\nbl + \nbr)/2m$ which is the velocity of the $c\bar{c}$ pair. Since the speed of the charm quark in $J/\Psi$ is
about 0.3\,c, we estimate the extra factor to be $\sim 0.3$ (and likely to be less). Dividing this factor from the $1^{-+}$ $D$ matrix elements
gives $\sim$ 0.20(2) which
is about a factor of two from that of the $2^{++}$ meson and comparable in size to the matrix elements of the other charmonium states.
Since the matrix elements of the lowest dimension operators (i.e. $D$-type) of the $1^{-+}$ in the charm region are comparable to and mostly smaller than those
of the other known charmonium states, it is not a hybrid by the criteria discussed in Sec.~\ref{sec:interpolation}.
On the other hand, the matrix element of $B$ for $1^{-+}$ is comparable to those of the other charmoniums, except $2^{++}$ which is
an order of magnitude smaller. This is presumably due to the factor of $\nblr/2m $ in the $2^{++}$ $B$ operator
in Table~\ref{tab:non-Oper}. Incorporating this factor of $\sim 0.3$ brings $B$ matrix element of $2^{++}$ to within a factor of 3 from that of the $1^{-+}$. In fact, all the P-wave operators have this $\nblr/2m$ factor and their matrix
elements will be comparable or larger than that of $1^{-+}$ when this factor is taken into account.
The fact that $1^{-+}$ does not have an extraordinarily large $B$ matrix element compared to other known charmonium states enhances the notion that it cannot be considered a hybrid in the charm region.


\subsection{Strange quark mesons}

    Next, we consider lighter quarkonium with the strange quark.
  The strange meson ($s\bar{s}$) masses in MeV are listed in Table~\ref{tab:smass}.

 \begin{table}[hbt]
\caption{Masses of strange quarkoniums from $\Gamma_w$- and $B_w$-type sources and point sinks.}
 \label{tab:smass}
 \begin{center}
 \begin{tabular}{c|cccc}
\hline
    &$ \Gamma_w\to \Gamma_p$&$ B_w\to \Gamma_p$&$ B_w\to B_p $&$ B_w\to D_p$\\
$0^{-+}$&$\ 714\pm\ 9$&$\ 750\pm\ 15$&$\ 713\pm\ 9   $&$\ 714\pm\ 10$   \\
$1^{--}$&$1027\pm\ 9 $&$1030\pm\ 12  $&$1030\pm\ 15   $&$1024\pm\ 12$   \\
$0^{++}$&$1570\pm63  $&$1566\pm21    $&$1568\pm21    $&$1567\pm21 $   \\
$1^{++}$&$1580\pm35  $&$1562\pm21    $&$1597\pm40    $&$1522\pm39$    \\
$1^{+-}$&$1613\pm35  $&$1569\pm18    $&$1608\pm54    $&$1598\pm19$   \\
$2^{++}$&--         &--           &$1638\pm21    $&$1611\pm60$    \\
$1^{-+}$&--         &--           &$2066\pm62    $&$2115\pm85$    \\
\hline
 \end{tabular}
 \end{center}
 \end{table}

\begin{figure}[h]
\begin{minipage}{0.5\linewidth}
{\includegraphics[scale=0.6]{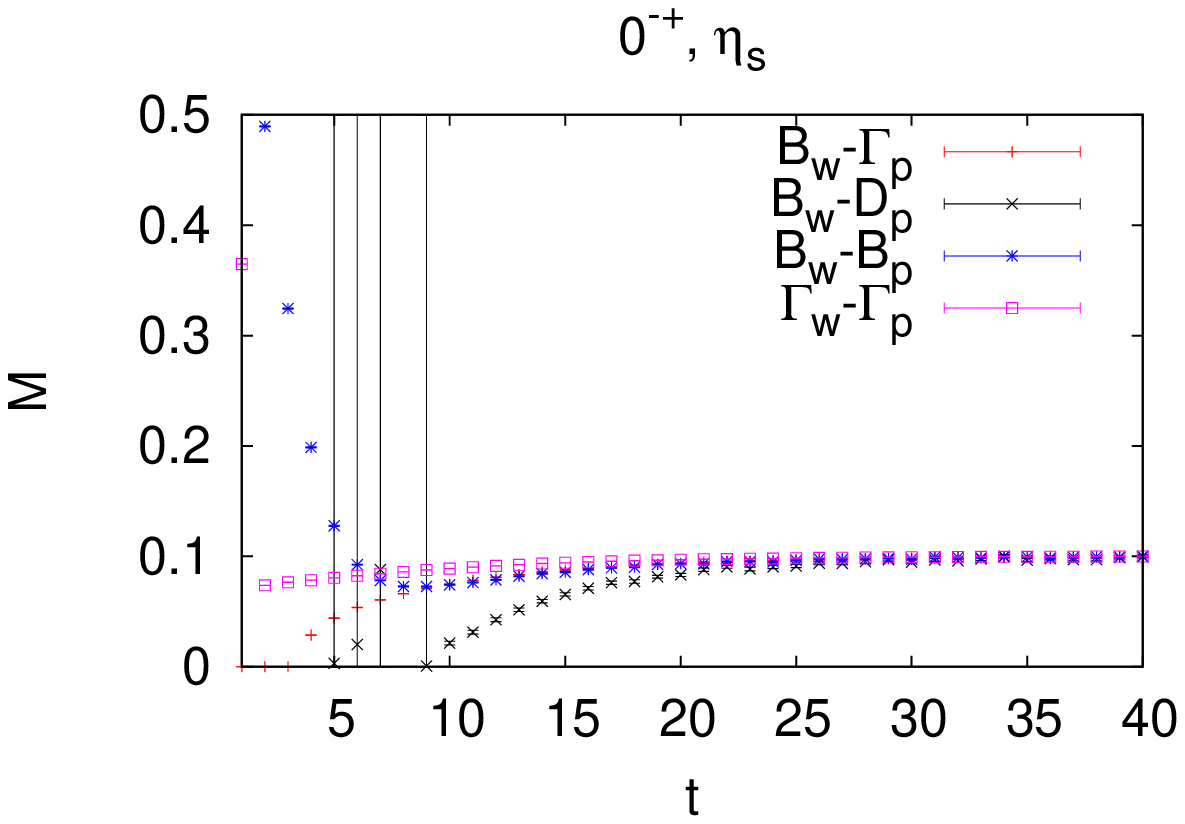}}
\end{minipage}%
\begin{minipage}{0.5\linewidth}
{\includegraphics[scale=0.6]{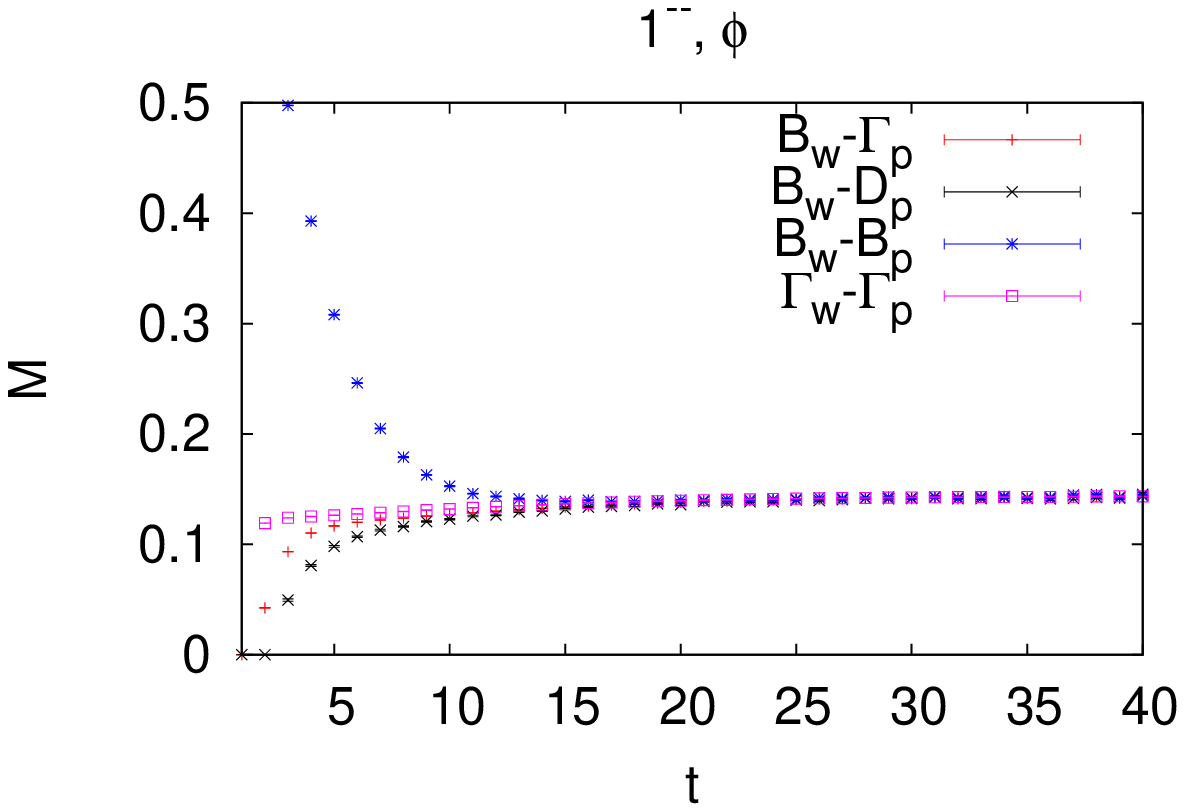}}
\end{minipage}%
\caption{\small Effective mass plot for $\eta_s$ and $\phi$ with $B_w$- and $\Gamma_w$-type wall sources
and $\Gamma_p, D_p$ and $B_p$ for the zero momentum point sinks.}\label{fig:swave_s}
\end{figure}

\begin{figure}[h]
\begin{minipage}{0.5\linewidth}
{\includegraphics[scale=0.6]{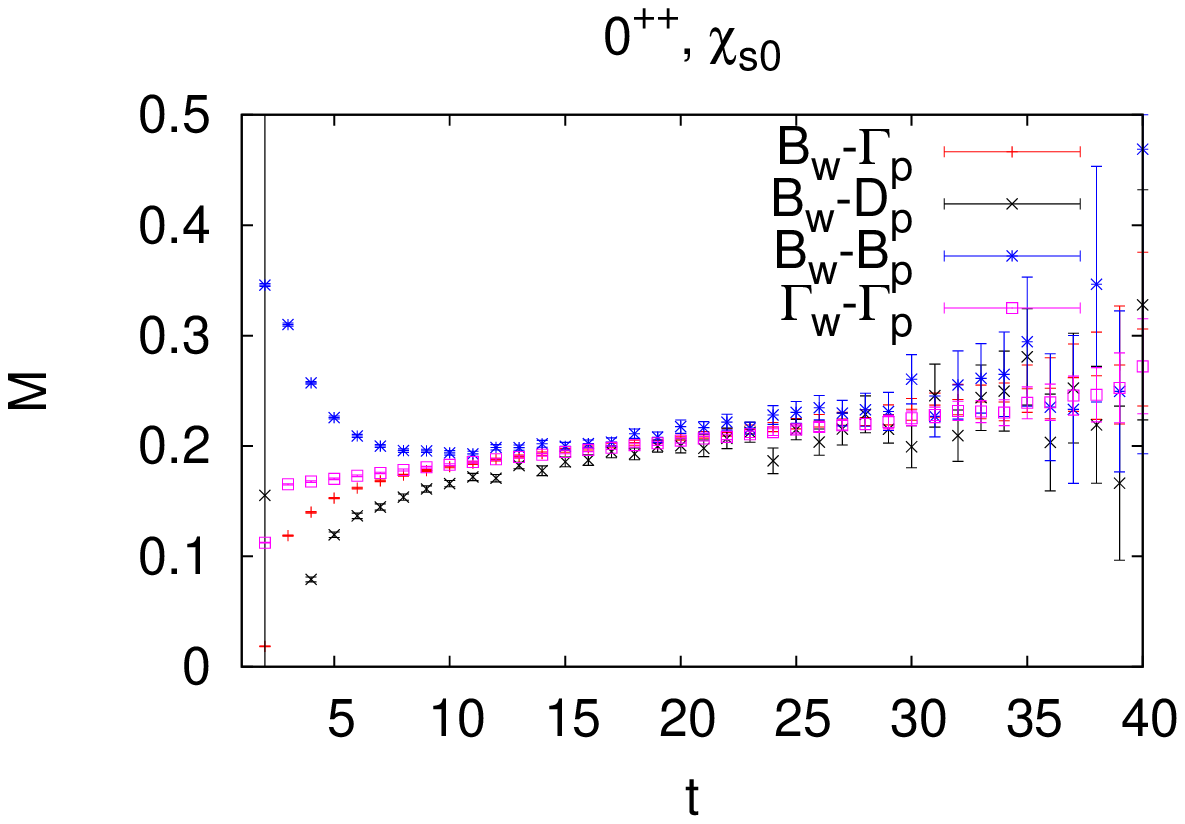}}
\end{minipage}%
\begin{minipage}{0.5\linewidth}
{\includegraphics[scale=0.6]{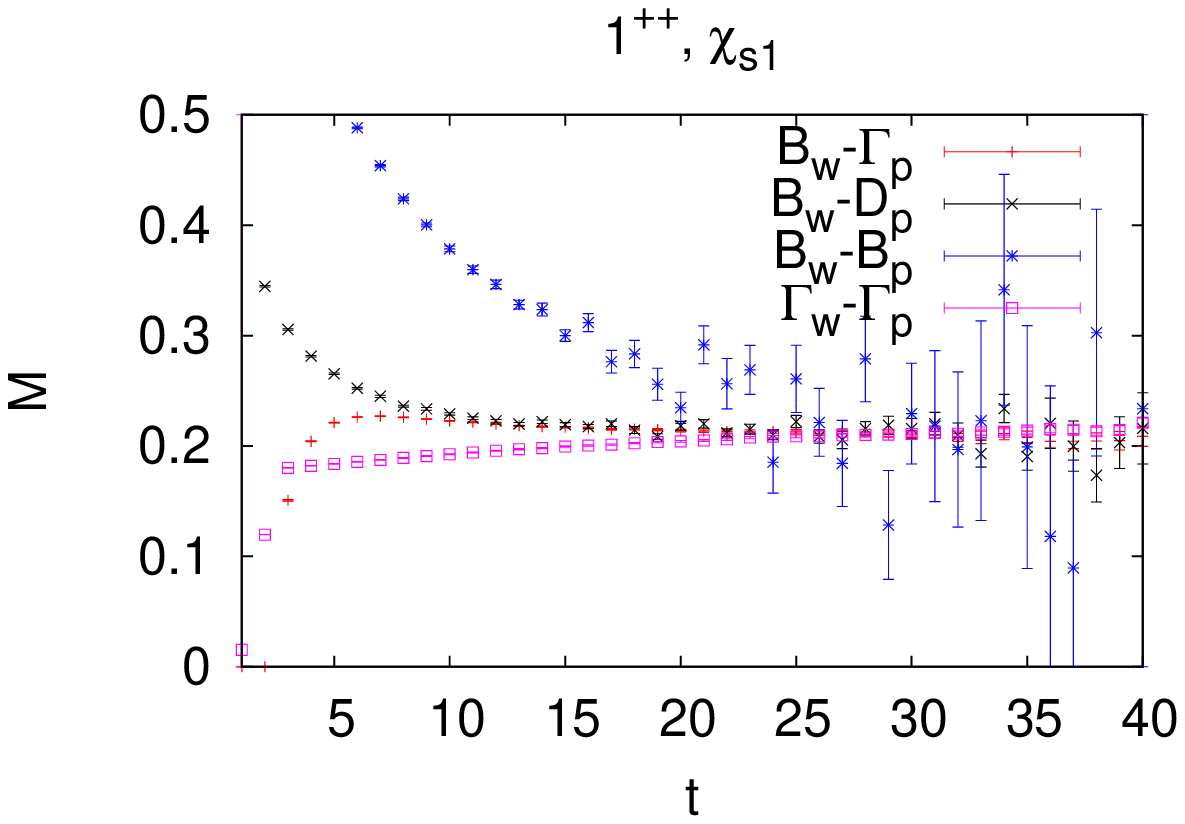}}
\end{minipage}
\caption{\small The same as Fig.~\protect\ref{fig:swave_s} for $f_{0(s)}$ and $f_{1(s)}$.}\label{fig:pwave1_s}
\end{figure}

\begin{figure}[h]
\begin{minipage}{0.5\linewidth}
{\includegraphics[scale=0.6]{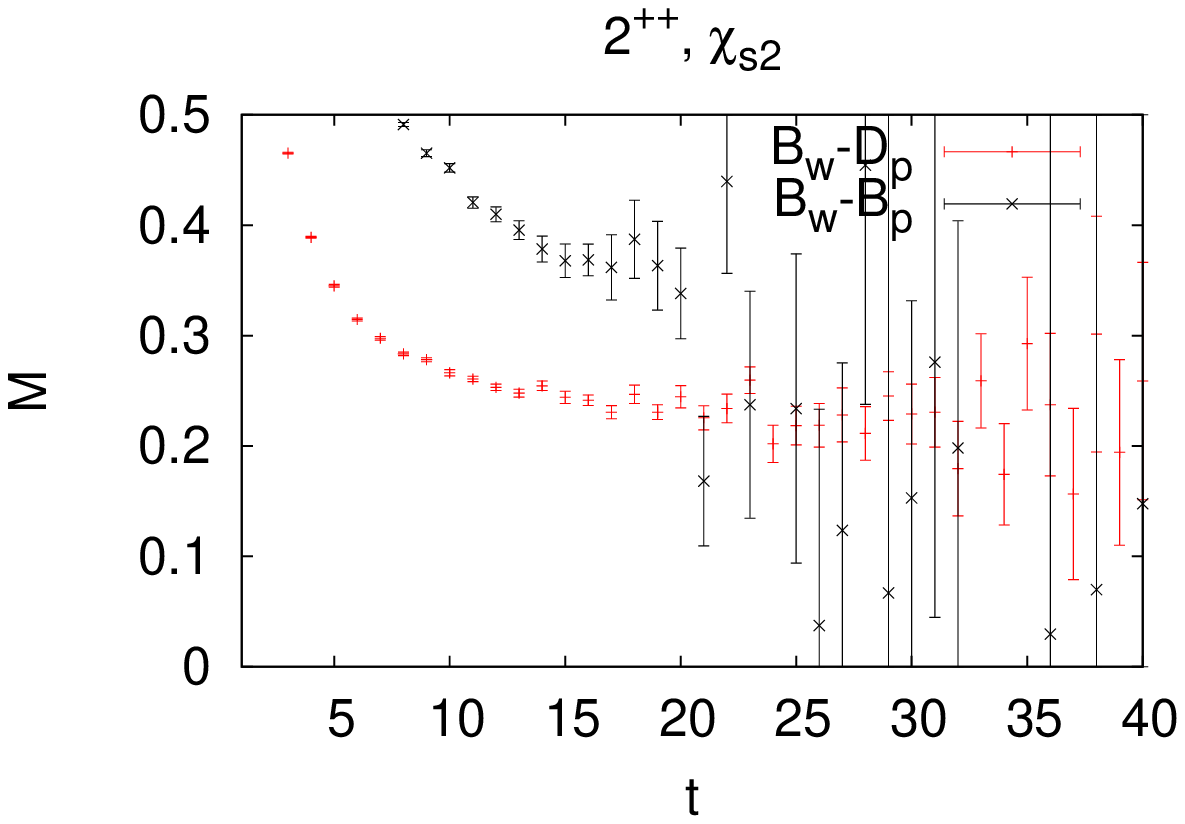}}
\end{minipage}%
\begin{minipage}{0.5\linewidth}
{\includegraphics[scale=0.6]{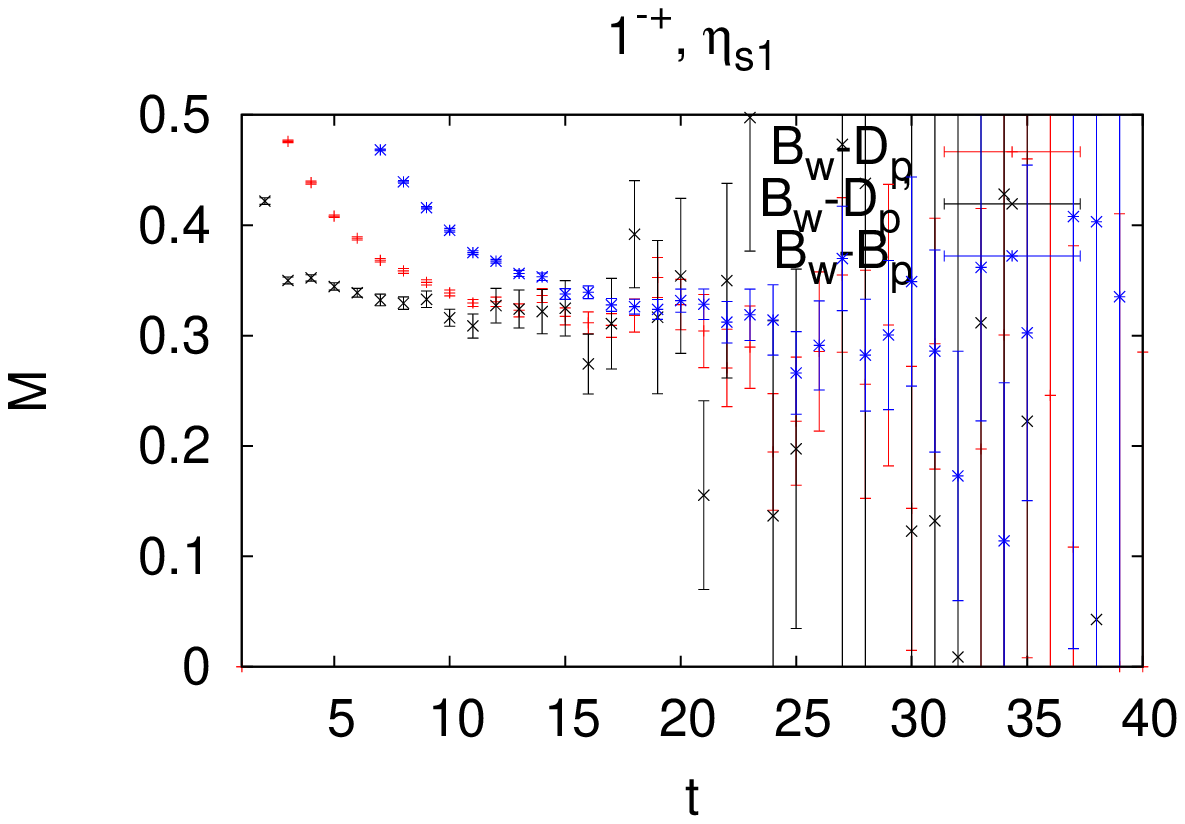}}
\end{minipage}%
\caption{\small The same as Fig.~\protect\ref{fig:swave_s} for $f_{2(s)}$ and $s\bar{s}$\,$ 1^{-+}$.}\label{fig:pwave2_s}
\end{figure}

The effective masses of $\eta_s (0^{-+}), \phi (1^{--}), f_{0(s)} (0^{++}), f_{1(s)} (1^{++}),
f_{2(s)} (2^{++})$ and the $s\bar{s}$\, $1^{-+}$ are plotted in Fig.~\ref{fig:swave_s}, \ref{fig:pwave1_s},
and  \ref{fig:pwave2_s}.

    We see from Table~\ref{tab:smass} and Figs.~\ref{fig:swave_s}, \ref{fig:pwave1_s},
and \ref{fig:pwave2_s} that the masses from different sources and sinks are the same within errors.
The matrix elements for $<0|\mathcal{O}_p|J^{PC}>$ for the $s\bar{s}$ mesons are listed in Table~\ref{tab:amp_strange}.

 \begin{table}[htb]
 \caption{The matrix elements  $<0|\mathcal{O}_p|J^{PC}>$ for strange quarkoniums.
 }\label{tab:amp_strange}
 \begin{center}
 \begin{tabular}{c|l|l|l}
\hline
    &$ \Gamma_p$ & $D$ & $B$ \\
\hline
$0^{-+}$  &$0.0247\pm0.0002$    &$0.021\ \pm0.002$   &$0.005\pm0.0001$  \\
$1^{--}$  &$0.0141\pm0.0002$    &$0.0113\pm0.0005$   &$0.0025\pm0.0001$  \\
$0^{++}$  &$0.043\ \pm0.006$    &$0.033\ \pm0.005$  &$0.017\ \ \pm0.004$\\
$1^{++}$  &$0.029\ \pm0.004$    &$0.034\ \pm0.004$  &$0.0018\pm0.0002$  \\
$1^{+-}$  &$0.019\ \pm0.006$    &$0.029\ \pm0.005$  &$0.0019\pm0.0004$ \\
$2^{++}$  &                     &$0.010\ \pm0.007$  &$0.0003\pm0.0001$\\
$1^{-+}$  &                     &$0.007\ \pm0.001$  &$0.004\pm0.001$  \\
          &                     &$0.006\ \pm0.002$ &\\
\hline
 \end{tabular}
 \end{center}
 \end{table}

 For the light quarkonium $s\bar{s}$, we do not expect the non-relativistic equivalence between the $D$-type and $B$-type operators to hold.
We shall compare the matrix elements directly. It is worthwhile noting that the dimension-4 $D$ matrix elements
 of $1^{-+}$ is comparable to that of the
$2^{++}$ meson and are not particularly smaller than those of the other $s\bar{s}$ mesons. Although the $B$
matrix element of $1^{-+}$ is larger than that of $2^{++}$, but it is not larger than those of other mesons. From these data, we see no evidence to distinguish the
$1^{-+}$ $s\bar{s}$ from other established $s\bar{s}$ mesons and identify it as a hybrid.

\section{Conclusion}\label{sec:conc}

   We set out to address the question: in view of the fact that there is a dimension-4
$\bar{\psi} \gamma_4  \nblr \psi$
interpolation field for $1^{-+}$, which does not involve the gauge field tensor, how does one identify it as a
hybrid and distinguish it from the ordinary mesons, which also have dimension-4 interpolation fields with a covariant derivative and dimension-5 interpolation fields involving explicitly the color
magnetic field $B$ in the form of $\epsilon_{ijk}\bar{\psi} \gamma_j \times B_k \psi$?  We emphasize that one cannot judge the nature of a state by the
appearance of its interpolation field. This is amply illustrated by the large matrix element $\langle 0|G\tilde{G}|\eta, \eta'\rangle$ which
shows that even though $\eta$ and $\eta'$ can be produced with the glue interpolation field, it does not mean that they are glueballs.
The glueball nature will be better revealed by a weak coupling to the $\bar{q}q$ interpolation field. We have also come up with
an example where the zero momentum operators of $\epsilon_{ijk}\bar{\psi} \gamma_j \times B_k \psi$ and $\bar{\psi} \gamma_4  \nblr \psi$
for the heavy quarks are the same up to a proportional constant, which is the quark mass. This implies that the former operator with a field tensor does not necessarily
project to an excited glue state, it could project to a state with the $q\bar{q}$ pair in a P-wave in the hadron as the latter interpolation field in the non-relativistic limit suggests.

    In light of this, we compare the matrix element of $\langle 0|\bar{\psi}\gamma_4\nblr \psi|1^{-+}\rangle$ and
$\langle 0|\bar{\psi}\eps\gamma_j\psi B_k|1^{-+}\rangle$ to the corresponding matrix elements of the other known $q\bar{q}$ mesons.
In the case of charmoniums, we find both the $D$- and $B$-type matrix elements of $1^{-+}$ are about the same size as the other mesons.
When a velocity of the $c\bar{c}$ pair is taken into account, they are also comparable to those of
$\chi_{c2} (2^{++})$, which is most similar to $1^{-+}$ in that neither has dimension-3 operator and their
dimension-4 operators are in the same Lorentz multiplet.
We have also examined the strange quarkoniums and found that the $D$- and $B$-type matrix elements
of $1^{-+}$ are comparable in size to those of the other $s\bar{s}$ mesons. Based on these data, we conclude
that there is not much distinction between $1^{-+}$ and other known $q\bar{q}$ mesons. There is no
evidence for it to be a hybrid.

    The leading non-relativistic expansion reveals that the dimension-4 operator $1^{-+}$ takes the form of
$\chi^{\dagger}\frac{1}{2m_c}(\sigma\cdot\nbl\nblr_i+\nblr_i\sigma\cdot\nbr)\phi$ and
$\chi^{\dagger}\frac{1}{2m_c}(\nbl_i\sigma_j\nblr_j+\sigma_j\nblr_j\nbr_i)\phi$. They involve a P-wave of the $q\bar{q}$ pair.
Since the center of mass of the $q\bar{q}$ in a constituent quark model is only a kinematical degree of freedom, confined
center- of-mass motion is not admissible in the constituent quark model. This is why the $J^{PC}$ of $1^{-+}$ and others involving the angular momentum of the $q\bar{q}$ pair are considered `exotic'.

In QCD, the $q\bar{q}$ pair can recoil against the non-excited glue field in the meson. Similarly, $q\bar{q}$ pair
can have orbital angular momentum relative to the bag in the MIT bag model, to the pion in the chiral quark model and to the flux-tube in the
flux-tube model. Thus in QCD and in models with additional constituents other than the $q\bar{q}$ pair, there can be
meson states with these `exotic' quantum numbers.
These additional $J^{PC}$ quantum numbers can be accommodated by supplanting the parity and angular momentum relations to
$P = (-)^{L + l + 1}$ and $\vec{J} = \vec{L} +\vec{l} + \vec{S}$.


\section*{Acknowledgments}
This work is partially support by U.S. DOE Grants No. DE-FG05-84ER40154 and the National Science Foundation
of China (NSFC) under the grant \# 11075167, 10835002, 10947007. We thank \mbox{B.A. Li} who pointed out to us his work on the $1^{-+}$ operator more than 30 years ago.



\begin{thebibliography}{99}

\bibitem{jj76}
R. Jaffe and K. Johnson, Phys. Lett. {\bf 60B}, 201 (1976).

\bibitem{djl81}
J. Donoghue, K. Johnson, and B.A. Li, Phys. Lett. {\bf 99B}, 416 (1981).

\bibitem{bcm82}
T. Barnes, F. Close, and S. Monaghan, Nucl. Phys. {\bf B198}, 380 (1982).

\bibitem{chp83}
C.E. Carlson, T.H. Hansson, and C. Peterson, Phys. Rev. {\bf D27}, 1556 (1983).

\bibitem{bar81}
T. Barnes, Z. Phys. {\bf C10}, 275 (1981).

\bibitem{cs82}
J. Cornwall, A. Soni, Phys. Lett. {\bf B120}, 431 (1983).

\bibitem{hm78}
D. Horn and J. Mandula, Phys. Rev. {\bf D17}, 898 (1978).

\bibitem{lw80}
K.F. Liu and C.W. Wong, Phys. Rev. {\bf D21}, 1350 (1980).

\bibitem{bar79}
T. Barnes, Nucl. Phys. {\bf B158}, 171 (1979).

\bibitem{bc82}
T. Barnes and F.E. Close, Phys. Lett. {\bf 116B}, 365 (1982).

\bibitem{cs83}
M. S. Chanowitz and S. R. Sharpe, Nucl. Phys. {\bf B222}, 211 (1983).

\bibitem{bcv83}
T. Barnes, F.E. Close, F. de Viron, and J. Weyers, Nucl. Phys. {\bf B224}, 241 (1983).

\bibitem{fps84}
M. Flensburg, C. Peterson, and L. Sk\"{o}ld, Z. Phys. {\bf C 22}, 293 (1984).

\bibitem{ip85}
N. Isgur and J. E. Paton, Phys. Rev. {\bf D31}, 2910 (1985).

\bibitem{bcs95}
T. Barnes, F.E. Close, and E.S. Swanson, Phys. Rev. {\bf D 52}, 5242 (1995).

\bibitem{lnp84}
J.I Latorre, S. Narison, P. Pascual, and R. Tarrach, Phys. Lett. {\bf B147}, 169 (1984).

\bibitem{cn00}
K.G. Chetyrkin and S. Narison, Phys. Lett. {\bf B485}, 145 (2000).

\bibitem{Huang:2011nv}
  P.~-Z.~Huang and S.~-L.~Zhu,

\bibitem{kk09}
Hyun-Chul Kim and Youngman Kim, JHEP {\bf 0901}, 034 (2009), [arXiv:0811.0645 [hep-ph]].

\bibitem{lmb97}
P. Lacock, C. Michael, P. Boyle, and P. Rowland, Phys. Rev. {\bf D 54}, 6997 (1996);
Phys. Lett. {\bf B401}, 308 (1997).

\bibitem{bhd97}
C. Bernard, {\it et al.}, Phys. Rev. {\bf D 56}, 7039 (1997).

\bibitem{jkm99} 	
K.J. Juge, J. Kuti, and C.J. Morningstar, Phys. Rev. Lett. {\bf 82}, 4400 (1999),
[hep-ph/9902336].

\bibitem{lm02}
X. Liao and T. Manke, [hep-lat/0210030].

\bibitem{bbg03}
C. Bernard,  {\it et al.}, Phys. Rev. {\bf D68}, 074505 (2003).

\bibitem{ml03}
Z.H. Mei and X.Q. Luo, Int. J. Mod. Phys. {\bf A 18}, 5713 (2003), [hep-lat/0206012].

\bibitem{hlw04}	
J.N. Hedditch, D.B. Leinweber, A.G. Williams (Adelaide U.), and J.M. Zanotti,
Nucl. Phys. Proc. Suppl. {\bf 129}, 248 (2004).

\bibitem{dep09}
J.J. Dudek, R.G. Edwards, M.J. Peardon, D.G. Richards, and C.E. Thomas,
Phys. Rev. Lett. {\bf103}, 262001 (2009);  Phys. Rev. {\bf D 82}, 034508 (2010).

\bibitem{dud11}
J. J. Dudek, Phys. Rev. {\bf D 84}, 074023 (2011), [arXiv:1106.5515 [hep-ph]].

\bibitem{ald88}
D. Alde {\it et el.}, Phys. Lett. {\bf B 205}, 397 (1988); D. R. Thompson {\it et al.},
Phys. Rev. Lett. {\bf 79}, 1630 (1997); S.U. Chung, {\it et al.}, Phys. Rev. {\bf D 60},
092001 (1999); A. Abele {\it et al.}, Phys. Lett. {\bf B 423}, 175 (1998);
A. Abele {\it et el., ibid.} {\bf B 446}, 349 (1999).

\bibitem{ada98}
G.S. Adam  {\it et el.}, Phys. Rev. Lett. {\bf 81}, 5760 (1998);
E.L. Ivanov {\it et el., ibid.}  {\bf 86}, 3977 (2001).

\bibitem{li75}
B. A. Li, Acta Physica Sinica, {\bf 24}, 21 (1975).

\bibitem{fks98}
T. Feldmann, P. Kroll, and B. Stech, Phys. Rev. {\bf D 58}, 114006 (1998); Phys. Lett. {\bf B 449}, 339 (1999).

\bibitem{cll09}
H.Y. Cheng, H.N. Li, and K.F. Liu, Phys. Rev. {\bf D 79}, 014024 (2009), [arXiv:0811.2577].

\bibitem{liu08}
K.F. Liu, `Scadron 70', AIP Conf. Proc. {\bf 1030}, 305 (2008), [arXiv:0805.3364 [hep-lat]].

\bibitem{reb75}
C. Rebbi, Phys. Rev. {\bf D 12}, 2407 (1975); {\it ibid.} {\bf D 14}, 2362 (1976).

\bibitem{fwt50}
L.L. Foldy and S.A. Wouthuysen, Phys. Rev. {\bf 78}, 29 (1950);
S. Tani, Prog. Theor. Phys. {\bf 6}, 267 (1951).

\bibitem{dj76}	
T.A. DeGrand and R.L. Jaffe, Annals Phys. {\bf 100}, 425 (1976).

\bibitem{lw82}
K.F. Liu and C.W. Wong, Phys. Lett. {\bf 113}, 1 (1982).


\bibitem{Klassen:1998ua}
  T.~R.~Klassen,
  Nucl.\ Phys.\  B {\bf 533}, 557 (1998)
  [arXiv:hep-lat/9803010].

\bibitem{Su:2004sc}
  S.~Q.~Su, L.~m.~Liu, X.~Li and C.~Liu,
  Int.\ J.\ Mod.\ Phys.\  A {\bf 21}, 1015 (2006)
  [arXiv:hep-lat/0412034].

\bibitem{tac06}
S. Tamhankar, A. Alexandru, Y. Chen, S.J. Dong, T. Draper, I. Horvath, F.X. Lee, K.F. Liu, N. Mathur, and J.B. Zhang (chiQCD collaboration),
Phys. Lett. {\bf B 638}, 55 (2006).




\end{thebibliography}
\end{document}